\documentclass[11pt,a4paper]{article}

\usepackage[utf8]{inputenc}
\usepackage[T1]{fontenc}
\usepackage{graphicx}
\usepackage{amsmath}
\usepackage{amsfonts}
\usepackage{amssymb}
\usepackage{amstext}
\usepackage{natbib}
\usepackage{makecell}

\usepackage[hmargin=0.7in,vmargin={1.2in,1.1in}]{geometry}
\usepackage{csquotes}
\usepackage{rotating}
\usepackage{booktabs}
\usepackage{csquotes}
\usepackage{amsbsy}
\usepackage{eqnarray}
\usepackage{lipsum}
\usepackage[affil-it]{authblk}
\usepackage{float}
\usepackage{setspace}
\usepackage{tabularx}
\usepackage{multirow}
\usepackage{color}
\usepackage{url}
\usepackage{soul}
\usepackage[official]{eurosym}
\usepackage[colorlinks]{hyperref}
\usepackage[dvipsnames,table,xcdraw,svgnames]{xcolor}
\usepackage{subfig}

\newcommand{\dotr}{\mbox{$\boldsymbol{\cdot}$}}

\definecolor{deeppink}{rgb}{1.0, 0.08, 0.58}

\graphicspath{{Figures/}}

\makeatletter
\renewcommand\footnotesize{%
	\@setfontsize\footnotesize\@ixpt{11}%
	\abovedisplayskip 8\p@ \@plus2\p@ \@minus4\p@
	\abovedisplayshortskip \z@ \@plus\p@
	\belowdisplayshortskip 4\p@ \@plus2\p@ \@minus2\p@
	\def\@listi{\leftmargin\leftmargini
		\topsep 4\p@ \@plus2\p@ \@minus2\p@
		\parsep 2\p@ \@plus\p@ \@minus\p@
		\itemsep \parsep}%
	\belowdisplayskip \abovedisplayskip
}
\makeatother

\AtBeginDocument{%
\hypersetup{
citecolor=MidnightBlue,
linkcolor=Brown,   
urlcolor=MidnightBlue}
}

\title{Innovation in times of Covid-19}

\title{Innovation in times of Covid-19}
\author[1,2,3]{Torsten Heinrich \thanks{torsten.heinrich@wiwi.tu-chemnitz.de}}
\author[4,2,3]{Jangho Yang \thanks{j634yang@uwaterloo.ca}}
\affil[1]{Department of Economics and Business Administration, Chemnitz University of Technology}
\affil[2]{Institute of New Economic Thinking at the Oxford Martin School, University of Oxford}
\affil[3]{Oxford Martin Programme on Technological and Economic Change, University of Oxford}
\affil[4]{Management Sciences, Faculty of Engineering, University of Waterloo}

\begin{document}

\maketitle

\begin{abstract}
    Did the Covid-19 pandemic have an impact on innovation? Past economic disruptions, anecdotal evidence, and the previous literature suggest a decline with substantial differences between industries. We leverage USPTO patent application data to investigate and quantify the disturbance. We assess differences by field of technology (at the CPC subclass level) as well as the impact of direct and indirect relevance for the management of the pandemic. Direct Covid-19 relevance is identified from a keyword search of the patent application fulltexts; indirect Covid-19  relevance is derived from past CPC subclass to subclass citation patterns. We find that direct Covid-19 relevance is associated with a strong boost to the growth of the number of patent applications in the first year of the pandemic at the same order of magnitude (in percentage points) as the percentage of patents referencing Covid-19. We find no effect for indirect Covid-19 relevance, indicating a focus on applied research at the expense of more basic research. Fields of technology (CPC mainsections) have an additional significant impact, with, e.g., mainsections A (human necessities) and C (chemistry, metallurgy) having a strong performance.
\end{abstract}

\section{Introduction}

1918, as the world was reeling from the effects of World War I and a devastating influenza pandemic, a sudden decline in patent applications and grants - in the US at least - went mostly unnoticed \citep{Berkesetal20}. Did it harm the economic development of the world? Did it delay the recovery after the war? It is difficult to tell for sure in the absence of detailed data. But it is certain that these two events had a significant impact on the technological development of the world - the two world wars precipitated many designs of future civilian products (aircraft, computers, rocketry) and may have accelerated progress through investment and experience curves \citep{Lafondetal20}, while also disrupting the research - and indeed the lives - of many innovators and researchers. As for the pandemic - locally different effects on future patent output have been shown \citep{Berkesetal20}, hence a profound effect seems likely. 

Almost 100 years on, in late 2019 and early 2020, a new pandemic disrupted both economic and social life at a scale previously unknown to this generation. Again, the effects on research, development, and innovation may be profound, but have yet to be analyzed. What are the determining factors? Do  the effects differ from one field of technology to another? Are government interventions necessary and expedient and, if so, what kind of interventions? The present paper contributes to answering these questions.

We leverage the public patent application data\footnote{https://bulkdata.uspto.gov/} provided by the United States Patent and Trademark Office (USPTO) to investigate both the trends by field of technology and any effect Covid-19 may have in particular fields. Of course, there are caveats to working with patent application data: Patents are not the same as innovations. Inventors may decide against patenting an invention since patenting is costly and also requires the specifications to be made public. Some patents, in turn, may not represent actual innovations but patent trolling, non-feasible ideas,\footnote{The patent office will check for novelty, obviousness, prior art, etc., but not for feasibility. That is up to the patent owner.} minor changes, etc. Patent applications come with the additional caveat that not every patent application has to be granted and that there is a time delay between the patent application being filed and it being made public (see figure~\ref{fig:age-at-publication} in appendix~\ref{app:data}). However, patent applications afford us a very early window into the development of new technologies. In addition, patents are expensive, and so patent applications indicate that the applicant believed they had made a discovery that merited patenting in terms of both novelty and potential revenue. This is mostly free of strategic maneuvering, pretence, and public-relations endeavours that may bias other data sources on innovation. Once all patent applications filed at a certain time have been made public, they also provide a full view of the population of patent applications in the USA, not just a sample.

Several recent studies have examined the impact of Covid-19 on innovation \citep{brem2021implications, dahlke2021crisis, wen2022impacts, bachmann2022r, agarwal2022drives, guderian2021innovation}, but little attention has been paid to its direct impact on invention activities and studies often only cover very specific fields of research.\footnote{\citet{alexopoulos2021evaluating} study the impact of Covid-19 on patent applications. Nevertheless, they only consider innovations in artificial intelligence, and the period of consideration ends in June 2021. To the best of our knowledge, our paper is the first attempt to analyze the full population of USPTO patent applications in a timeframe with the potential to cover a significant part of patents applications filed in the time period after the onset of the pandemic in early 2020.} Our study fills this gap and provides empirical evidence on patenting in general during the Covid 19 pandemic. We identify Covid-19 relevant patents by performing keyword searches in the patent application fulltexts. According to our analysis, direct Covid-19 relevance has a strong, positive, and highly significant effect on the growth of patent applications during the first year of the pandemic, in contrast to an overall decline in patenting activity during that period. However, indirect Covid-19 relevant patents, which are derived from past CPC\footnote{CPC is the Cooperative Patent Classification system, a system of classifying patents and patent applications, that has been jointly developed by the European Patent Office and the USPTO and is in use since 2013 \cite{USPTO/EPO17}.} subclass to subclass citation patterns, are not particularly associated with the increase in the number of patent applications. Additionally, we find that there is significant variation in the baseline growth of patent applications across different fields. As an example, CPC mainsections A (human necessities) and C (chemistry, metallurgy) perform very well, whereas mainsections B (operations, transport) and F (engineering) perform less well.

Considering the serious efforts being made in the development of technology to aid and guide the response to Covid-19 - vaccines in particular \citep{Florio22} - the positive effect of direct Covid-19 relevance, that we identify here, is somewhat predictable. Nonetheless, it is not immediately apparent that it would be reflected in patent application numbers. This implies that Covid-19 relevant inventions were quickly developed and were patented and hence shared with competitors during the pandemic. Based on the full public USPTO patent application data for the timeframe, this paper reconsiders hypotheses from anecdotal and fragmented discussions on how the pandemic affected innovation. It provides potential avenues to fine-tune innovation policy so that post-pandemic research activities can be stimulated, particularly in only indirectly Covid-19 relevant fields (such as CPC classes C07 - organic chemistry, G06 - computing, or H04 - electric communication), which have suffered during the pandemic.

The following sections discuss the literature (section~\ref{sect:literature}) and our data sources and data preprocessing (section~\ref{sect:data}). Section~\ref{sect:empirical} presents our empirical approach and results and section~\ref{sect:conclusion} concludes.

\section{Literature}
\label{sect:literature}

\paragraph{Heterogeneous effects of the pandemic on patenting}

Early accounts from trade publications, e.g. \citet{Gaudry/Gianola20}, warn of a substantial decline in patent applications, especially in fields like chemistry and life sciences. At the same time, an unprecedented effort was under way to develop the vaccines and other pharmaceutical technology to manage, control, and prevent Covid-19 infections \citep{Sampat/Shadlen21}. Because of the short time span since the start of the pandemic, the number of detailed empirical papers on the topic is limited. Researchers either had to work with very incomplete patent application data sets \citep{Bloometal21,Yoon21} or they had to leverage other data sources \citep{Han/Qian20}, that come with problems of their own. The available research does, however, consistently highlight significant differences between different industry sectors (or fields of technology) in the development of innovation:

\citet{Bloometal21} find that the number of patent applications related to work-from-home technologies doubled in the first six months of 2020. They conduct a keyword search for terms related to remote work. Their findings are plausible considering the timeframe - a time when most employees in office jobs, but also teachers, community organizers, contract negotiators, and people in many other occupations had to work from home, and the economic potential of those technologies became evident. However, as the study acknowledges, coming just a few months into the pandemic and hence only being able to include patent applications published within a few months, it has to make due with an incomplete data set. This data may be subject to bias, e.g. because of eagerness of inventors in some fields to have their work in the public record (for signaling for acquiring further funding) or because of intentional or unintentional preferential processing of some fields by the patent office for a variety of reasons.

\citet{Han/Qian20}, analyzing iFinD data on Chinese public listed companies, find that R\&D investment has increased during the pandemic. This holds across all sectors, but they also identify substantial differences between industries with IT at the low end and finance at the high end. They conjecture that R\&D investment can be used as a proxy for R\&D capabilities, hence potentially future innovation. 

Working with data from the Korea Intellectual Property Rights Information Service (KIRPIS), \citet{Yoon21} finds that the number of patent applications in Korea has not significantly declined during the Covid-19 crisis. There were, however, significant differences between fields of technology with a surge in remote collaboration technology driven by innovations by small companies. As KIRPIS also only includes published patent applications and South Korean law defines the same 18-months-term that applies to the USA (see section~\ref{sect:data} below), this paper cannot have had access to the full (or even an almost complete) population of South Korean patent applications though.

\paragraph{Disruptions of innovation}

Another literature tradition considers disruptions of technological development in general, e.g. as a result of economic crises \citep{yamashita2021economic, brem2020impact, archibugi2013economic, amore2015companies} or standardization events like the emergence of patent pools \citep{kim2016impact}. The overall effect of economic crises on innovation is unsurprisingly negative \citep{yamashita2021economic}. However, there are secondary effects to the innovation system: consolidation of designs \citep{brem2020impact} as well as concentration of innovation to fewer highly-innovative firms \citep{archibugi2013economic} and to those that have built resilience conducting R\&D in past crises \citep{amore2015companies}. It remains to be seen, if similar patterns emerge during and in the aftermath of the Covid-19 pandemic.

\paragraph{Innovation policy response to the pandemic}

Finally, another important line of research assesses the innovation policy response to the Covid-19 pandemic. In a Science editorial written in the early days of the pandemic, \citet{Zerhounietal20} suggest the creation of milestone-based subsidies to expedite the development of Covid-19 related technologies - drugs, diagnostics, vaccines. In a conceptual paper drawing on earlier empirical results, \citet{Gross/Sampat21} consider innovation policy in normal and crisis times. They take World War II and the Covid-19 pandemic as examples for crises and look at the policies and funding programs of the World-War-II-time U.S. Office of Scientific Research and Development (OSRD) in particular. Starting from Arrow's \citeyearpar{Arrow62} and Nelson's \citeyearpar{Nelson59} principal features of innovation in normal times (indivisibility, inappropriability, uncertainty), they find that crisis times' need for quick results suggests prioritizing applied research, encouraging redundancy (e.g., funding multiple vaccines), and loosening appropriability guidelines. While the Covid-19-pandemic did not see anything like the OSRD, there were aggressive public funding programmes in the USA and multiple other countries. \citet{Gross/Sampat21} do not make predictions regarding the development of innovation, but their analysis would suggest an acceleration of innovation (and patenting) in fields related to Covid-19 treatment and vaccine deployment etc. with a focus on practical applications over further theoretical groundwork.

Florio \citeyearpar{Florio21, Florio22} also finds that Covid-19 related innovations were funded publically to a substantial part. What is more, the immediate development of treatments and vaccines is based on overwhelmingly publically funded basic research. His analysis concludes, however, that in the presence of subsidies at this scale, patents as instruments for the protection of intellectual property rights may not be necessary. At least one open-source vaccine is in existence.


\section{Data}
\label{sect:data}

\subsection{Data sources}

To assess the effect of the pandemic on innovation we use patent application data from the first year of the Covid-19 pandemic, i.e. the year starting when the pandemic was first commonly recognized as a major global public health crisis and when nearly all governments including the United States introduced severe containment measures such as stay-at-home orders, work-from-home orders, closure of entertainment venues, restrictions on travel, etc. This is the year from April 2020 to March 2021. We contrast this period to the previous years, in various analyses the year immediately preceding it (April 2019 - March 2020), and average data across the years from April 2014 through March 2020. We make the simplifying assumption that the Covid-19 pandemic was the single dominant event occurring in this period and the reason precipitating basically all empirical changes between the years preceding the pandemic and the first year of the pandemic.

We use patent application bulk data\footnote{https://bulkdata.uspto.gov/} provided by the USPTO (United States Patent and Trademark Office). Following 35 U.S. Code Section 122 all patent applications have to be published within 18 months of being filed, except for specific exceptions (subject to confidentiality, design patent applications, withdrawn patent applications). We only consider patent applications that were published within 18 months (specifically 555 days\footnote{One and a half years plus seven days, since patent applications are published only once a week.}), both for patent applications during and before the pandemic for consistency. This constitutes the bulk of all patent applications; of the patent applications published in the relevant period, a majority of $76.67\%$ were published within 18 months (555 days) after being filed (see figure~\ref{fig:age-at-publication} in Appendix~\ref{app:data}). With this limit, we also have a complete picture of the patent applications in the first year of the pandemic as of October 2022. 


\subsection{Preprocessing}

\begin{figure}[H]
	\begin{center}
 \includegraphics[width=.7\textwidth]{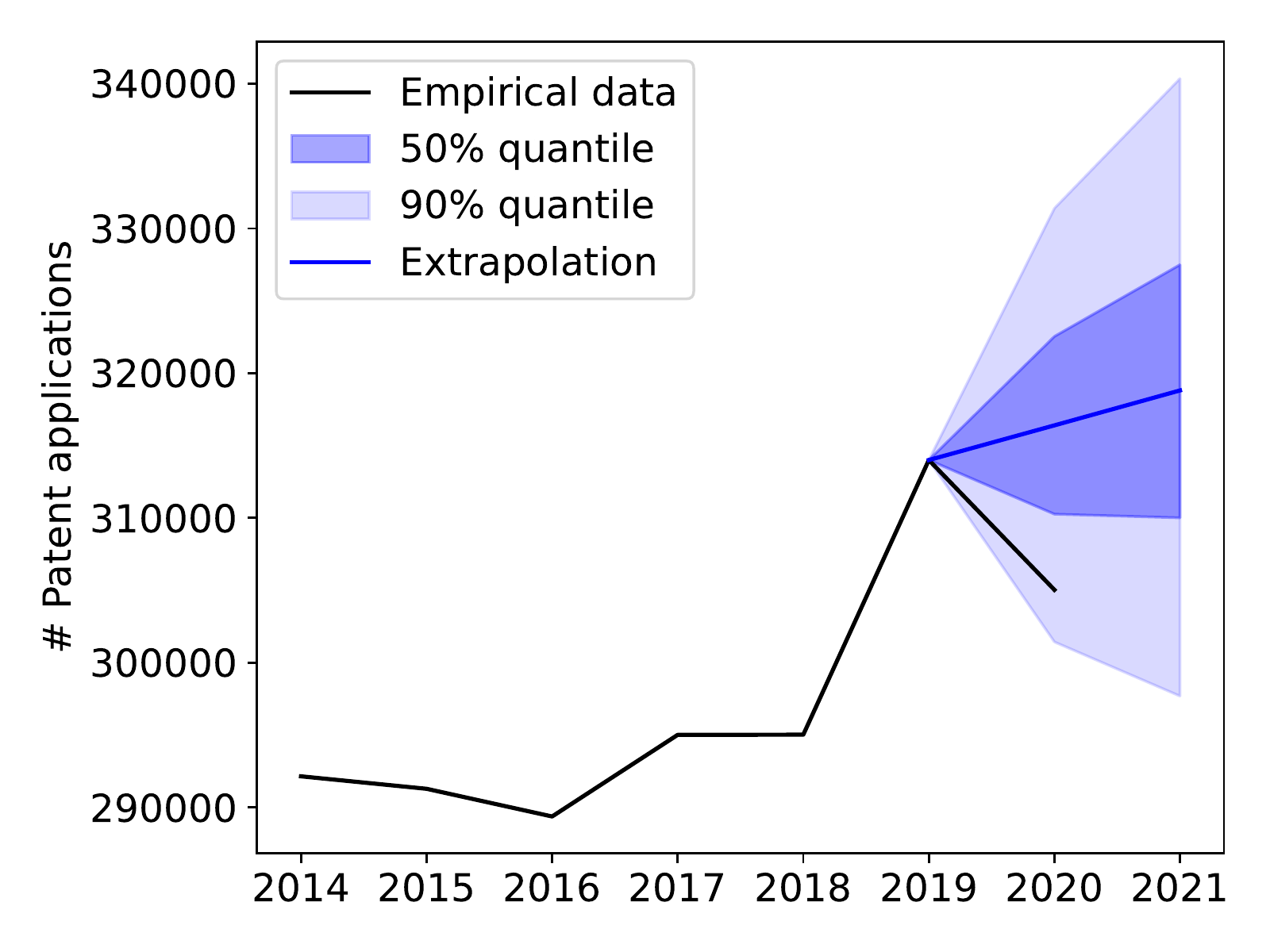}
\caption{Patent applications by year and mainsection, extrapolation and realization for 2020 and 2021.}
\label{fig:growth-all}  
	\end{center}
\end{figure}

\begin{figure}[H]
	\begin{center}
      \includegraphics[width=.7\textwidth]{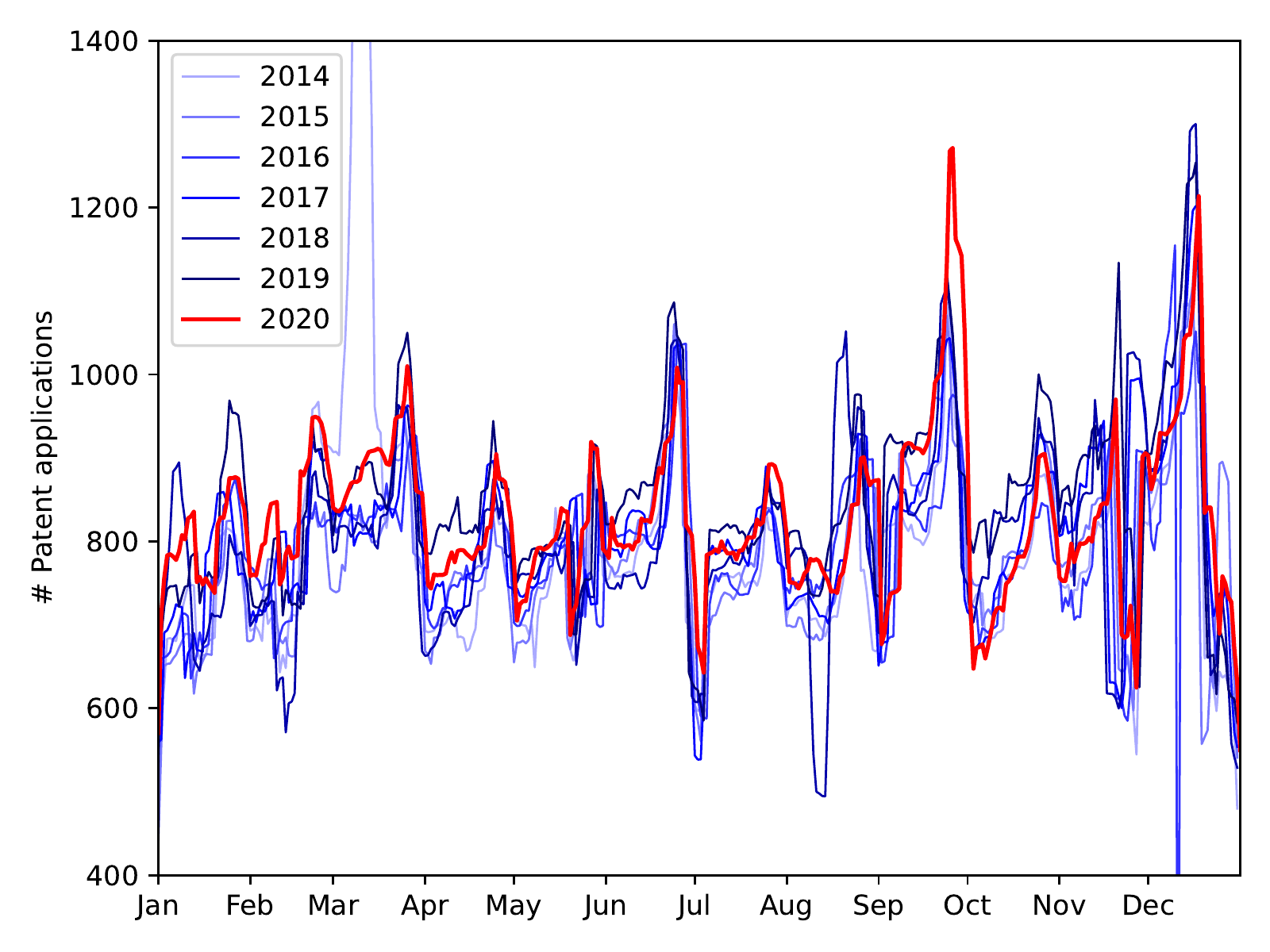}
\caption{Number of patent applications filed by day and year 2014-2020, that were published within 18 months.}
\label{fig:base-data}  
	\end{center}
\end{figure}

In total, we have 2,079,876 patent applications between April 2014 and March 2021 that were published within 18 months. Figures~\ref{fig:growth-all} and \ref{fig:base-data} show the number of patent applications by year and by day (corrected for the weekly cycle with a moving average filter) between 2014 and 2020. It is clear that the increasing trend has ceased and the total number has gone down slightly with the pandemic, but with substantial variation over the year - and also across fields, as shown below. 

To assess the effect for specific fields, we aggregate the number of patents to the level of CPC subclasses \citep{USPTO/EPO17} as listed on the patent application and weighted with the total number of subclasses on a patent. I.e., a patent application with the following three codes (two of which belong to the same subclass A61K), A61K 9/50 (Preparations for medical purposes - Microcapsules)\footnote{The descriptions are shortened for clarity; the full descriptions can be found on for instance on the website of the USPTO, \href{https://www.uspto.gov/web/patents/classification/cpc/html/}{https://www.uspto.gov/web/patents/classification/cpc/html/}.}, A61K 47/68 (Medicinal preparations - modifying agents chemically bound to the active ingredient - antibody or immunoglobulin), and C12N 5/00 (Microorganisms or enzymes - undifferentiated human, animal or plant cells) is counted to the two subclasses A61K and C12N with 50\% weight each.\footnote{The weights are evenly split between the subclasses the patent application belongs to and sum to one. E.g., a patent application with 5 subclasses counts with 20\% to each of those. Multiple codes of the same subclass do not shift the weights.}

\subsection{Variable construction}
\paragraph{Direct Covid-19 relevance}

To identify the patent applications that are directly relevant for the response to the Covid-19 pandemic, we perform a keyword search in the full text of the patent applications for the following keywords:
\begin{itemize}
  \item ``Covid-19''
  \item ``Covid19''
  \item ``2019-nCov''
  \item ``SARS-CoV-2''
  \item ``Pandemic'' and ``Coronavirus''
  \item ``Pandemic'' and ``Covid''  
\end{itemize}
The text is transformed to lower case for the keyword search, so different capitalization conventions will have no effect. 

We then compute the share of directly Covid-19 relevant patent applications per subclass, yielding the vector $CovidRelevant$ that indicates direct Covid-relevance by CPC subclass. As above, patent applications $i$ assigned to $n_i$ subclasses are counted as $1/n_i$ for each of those subclasses, thus
\begin{equation}
CovidRelevant_j = \sum_i \frac{I(i,j)}{n_i} ,    
\end{equation}
where $I(i,j)$ is the indicator that patent application $i$ belongs to subclass $j$ (assuming value 1 if it does, 0 otherwise). A list of the 50 subclasses with the highest values of this variable can be found in table~\ref{tab:C19direct} in appendix~\ref{app:c19relevance}

We also define a binary variable that assigns value $1$ to the 50 subclasses with the highest continuous Covid-19 relevance variables ($CovidRelevant$) and value $0$ to all others:
$$BinaryCovidRelevant_i = \begin{cases}
  1 \quad\quad \textup{if } D(CovidRelevant_i)\\
  0 \quad\quad \textup{otherwise}
\end{cases}$$ where $D$ is the indicator function, indicating that the subclass is among the 50 with the highest $CovidRelevant$ values.

We do this to complement our regressions, that use continuous Covid-19 relevance, with equivalent ones with a binary variable in section~\ref{sect:robustnesscheck}. Those with binary regressors make the scale of the impact of the coefficient more illustrative and easier to interpret. E.g., predicted growth for Covid-19 relevant subclasses is positive, if the sum of the regression intercept and the coefficient corresponding to $CovidRelevant$ is positive, unless other continuous variables are present and significant.\footnote{We shall see that the contributions of other continuous variables will be relatively minor in scale.} For mainsection fixed effects, their coefficients can be added.\\

\paragraph{Indirect Covid-19 relevance}

In addition, we use the full subclass to subclass citation frequency matrix $\mathbb{C}$ computed from USPTO patent data to assess which subclasses are indirectly relevant for Covid-19. Matrix $\mathbb{C}$ is defined as follows: An entry $c_{i,j}$ in row $i$, column $j$ is the share of the backward citations from subclass $i$ patents that goes to subclass $j$, where multiple citations from the same subclass to the same subclass on the same patent are counted as only one citation.\footnote{Multiple such citations may occur, because there may be citations to different finer classification levels that belong to the same subclass. We choose to count them only once, as they indicate utilization of very similar types of prior knowledge for the new innovation the patent represents.} The vector of indirect Covid-19 relevance by subclass is simply computed as:
\begin{equation}
\label{eq:IndirectCovidRelevant}
  IndirectCovidRelevant = CovidRelevant^T \quad \times \quad \mathbb{C}.
\end{equation}
where $^T$ stands for the transpose. Equation~\ref{eq:IndirectCovidRelevant} multiplies a $(1\times m)$-matrix (direct Covid-19 relevance vector transposed) with a $(m\times m)$-matrix (subclass-to-subclass citation matrix) to yield a $(m\times 1)$-matrix (indirect Covid-19 relevance vector). 
A list of the 50 subclasses with the highest values of this variable can be found in table~\ref{tab:C19indirect} in appendix~\ref{app:c19relevance}.

\section{Empirical results}
\label{sect:empirical}

\subsection{Empirical method}
\label{sect:model}

We aggregate the values to the level of CPC subclasses (667 in total) to gain more statistical power. The dependant variable we are interested in is the annual growth of the number of patent applications (by subclass) in the year after the start of the most severe phase of the pandemic and the pandemic response measures in most countries: the year from April 2020 to March 2021 (compared to April 2019 to March 2020). We consider the following regression model:

  \begin{equation}
    \label{eq:model7}
    \begin{array}{lll}
    Growth2020 = \beta_0 & + \beta_1 \log(NumberPreCovid) + \beta_2 PastGrowth + \beta_3 CovidRelevant  & \\
                         & + \beta_4 IndirectCovidRelevant + MainsectionFE + \varepsilon
    \end{array}
  \end{equation}

For $PastGrowth$, we consider the annual growth in the year before the pandemic (April 2019 to March 2020 compared to the period April 2018 to March 2019). However, we perform a robustness check, in which we use the average annual growth in the 5 years leading up to the pandemic (always April to March) instead.\footnote{Either of these two measures could plausibly have an effect on $Growth2020$, either a positive effect because of rigidity (autocorrelation) or a negative one because of saturation. However, the two measures would be expected to be strongly correlated (one featuring in the computation of the other), and should thus not be used in the same model but considered separately. Pearson correlation is $\varrho=0.41$, Spearman rank correlation is $r=-0.22$.}

We perform regressions both using ordinary least squares (OLS) fitting and a robust regression with Huber's norm \cite{Huber73} using the implementation from Python Statsmodels \citep{Seabold/Perktold10}. To assess the viability of the OLS regression, we consider the residual skew and kurtosis as well as the Jarque-Bera statistic \citep{jarque1980efficient}, all of which can be used to test for the normality of the residuals. For the robust regression, we report the following goodness statistics\footnote{The usual (adjusted) R$^2$ is not a good metric for the goodness of robust regressions since it is can also be dominated by the tails of the distribution, just like the OLS fit.}: 
\begin{itemize}
  \item Root mean square error $RMSE = \sqrt{\mathbb{E}(\hat{y}-y)}$, an extension of the standard deviation of the estimator to biased samples.
  \item Mean absolute deviation $MAD = \mathbb{E}(|\hat{y}-y|)$
  \item Median absolute deviation $MAD = \mathbb{E}(|\textup{Median}(y)-y|)$
  \item Error inter-quartile range $EIQR = QF(0.75)-QF(0.25)$, where $QF$ is the quantile function (inverse of the cumulative distribution function) of the sample of $\hat{y}-y$.
\end{itemize}
All these measures should be as small as possible for a good fit. For our purposes, it is not problematic that the measures are not normalized, as our regressions are performed on almost the same samples (except for a handful cases in which we have some missing observations). 

\subsection{Primary results}
\label{sect:results}

\paragraph{OLS fit leads to non-normal residuals}

The OLS regressions in tables~\ref{tab:ols:g19} shows highly significant and positive coefficients for Covid-19 relevance in all seven models considered. Indirect Covid-19-relevance may be significant in model (7), but barely, with a p-value of $9.5\%$. The control variables considered are also found to be significant in most cases, $\log(NumberPreCovid)$ with a negative coefficient, $PastGrowth$ with varying signs and varying degrees of significance.

\begin{table}[H]
	\footnotesize

\begin{tabular}{p{3.3cm}ccccccc}
  \hline
                          & (1)           & (2)           & (3)           & (4)           & (5)           & (6)            & (7)            \\ 
  \hline
(Intercept)                  & -0.0415$^{**}$ &  0.1149$^{**}$  &   0.1584$^{**}$  &    0.1001$^{*}$   &     0.1092$^{*}$   &      0.1469$^{*}$   &       0.2463$^{***}$ \\        
                            &  (0.0135)   &   (0.0427)   &    (0.0587)   &     (0.0453)   &      (0.0433)   &       (0.0587)   &        (0.0596)  \\         
log(Number pre-Covid)        &             &  -0.0257$^{***}$ &   -0.0284$^{***}$ &    -0.0238$^{**}$ &     -0.0250$^{***}$ &      -0.0263$^{***}$ &       -0.0418$^{***}$\\        
                            &             &   (0.0067)   &    (0.0071)   &     (0.0075)   &      (0.0067)   &       (0.0072)   &        (0.0079)  \\         
Growth 2019                  &             &              &               &                &     0.0134$^{***}$  &      0.0133$^{***}$  &       -0.2265$^{***}$\\        
                            &             &              &               &                &      (0.0035)   &       (0.0035)   &        (0.0280)  \\         
Covid-relevant               & 75.3934$^{**}$ &  93.1178$^{***}$ &   95.9183$^{***}$ &    83.5170$^{**}$ &     84.5140$^{***}$ &      87.2576$^{***}$ &       101.7010$^{***}$\\        
                            &  (26.2604)  &   (26.3862)  &    (27.1556)  &     (26.2676)  &      (26.2213)  &       (27.0070)  &        (25.5949) \\         
Indirectly Covid-relevant    &             &              &               &    8.5799      &                 &                  &       87.8901$^{\dotr}$  \\        
                            &             &              &               &     (51.2511)  &                 &                  &        (52.5284) \\         
   \hline
Mainsection FE            &               &               &  O            &               &               &  O              &  O              \\
   \hline
N                         &  645          & 645           & 645           & 643           & 641           & 641            & 640            \\
df                        &  643          & 642           & 634           & 639           & 637           & 629            & 627            \\
Adj R$^2$                 &  0.0111     &   0.0320     &    0.0361     &     0.0243     &      0.0511     &       0.0556     &        0.1246    \\         
p-value (F-Statistic)     &  0.004        & 0.000         & 0.000         & 0.000         & 0.000           & 0.000          & 0.000            \\
AIC                       &  414.8        & 402.0         & 407.2         & 380.7         & 376.9         & 381.8          & 310.8           \\
Jarque-Bera               & 3.5$\times 10^4$ & 3.0$\times 10^4$ & 3.0$\times 10^4$ & 3.4$\times 10^4$ & 3.5$\times 10^4$ &  3.5$\times 10^4$ & 3.1$\times 10^4$\\ 
Residual skew             &  4.3          & 3.8           & 3.8           & 3.9           & 4.0           & 4.0            & 3.7            \\
Residual kurtosis         &  38.3         & 35.3          & 35.7          & 37.6          & 38.1          & 38.2           & 36.1           \\
   \hline
  \multicolumn{7}{c}{Signif. codes:  $^{***}$ $p<0.001$, $^{**}$ $p<0.01$, $^{*}$ $p<0.05$, $^{\dotr}$ $p<0.1$}\\
\end{tabular}
\caption{OLS regression with growth 2019 (in the year before the pandemic, i.e. 04/2019-03/2020) as PastGrowth.}
\label{tab:ols:g19}
\end{table}
  
However, the normality tests for the residuals (Jarque-Bera statistic, kurtosis, and skew\footnote{For normality, the Jarque-Bera statistic should be low, skew should be near 0, and kurtosis should be 3.}) consistently show that the residuals are not Gaussian. The OLS fit is therefore most likely invalid and possibly misleading. The distribution of the residuals in model (7) is shown in figure~\ref{fig:residuals:lin} in appendix~\ref{app:ols_res}; it is clear that the non-normality is due to both outliers (or heavy tails), left-skewedness, and high kutrosis. We therefore perform the fit again using a robust regression with Huber's T norm \cite{Huber73}. Alternative approaches, such as exclusion of outliers, would not only require manipulating the data, they would also fail to address other problems such as the skew and the kurtosis.\\

\paragraph{Robust regression fit}

The robust regressions in tables~\ref{tab:rob:g19} shows strong significance for direct Covid-relevance (at the $99.9\%$ level) with a positive coefficient for all seven models. Evidently, the (directly) Covid-relevant fields of research saw a substantially better growth performance in the first year of the pandemic compared to other fields of inquiry. E.g., in model (6), an increase of the share of the patents addressing Covid-19 in a subclass by $1$ percentage point is associated with an increase of the growth rate in 2020 by $0.706$ percentage points.\footnote{I.e. $70.6229$ growth percentage points for $100$ percentage points in Covid-19 relevance, corresponding to the coefficient of $70.6229$ in table\ref{tab:rob:g19}.} This raises the baseline expectation to positive growth with direct Covid-19 relevance. We will return to this point in section~\ref{sect:robustnesscheck}.

\begin{table}[H]
\footnotesize
\begin{tabular}{lccccccc}
  \hline
                          & (1)           & (2)           & (3)              & (4)           & (5)           & (6)           & (7)          \\ 
  \hline
(Intercept)                  & -0.0734$^{***}$ &  -0.0648$^{**}$ &   -0.0022     &    -0.0916$^{***}$ &     -0.0683$^{**}$ &      -0.0096     &       0.0421     \\        
                            &  (0.0075)   &   (0.0239)   &    (0.0323)   &     (0.0257)   &      (0.0243)   &       (0.0324)   &        (0.0349)  \\         
log(Number pre-Covid)        &             &  -0.0013     &   -0.0052     &    0.0034      &     -0.0009     &      -0.0041     &       -0.0112$^{*}$ \\        
                            &             &   (0.0037)   &    (0.0039)   &     (0.0043)   &      (0.0038)   &       (0.0040)   &        (0.0046)  \\         
Growth 2019                  &             &              &               &                &     0.0166$^{***}$  &      0.0163$^{***}$  &       -0.1413$^{***}$\\        
                            &             &              &               &                &      (0.0020)   &       (0.0019)   &        (0.0164)  \\         
Covid-relevant               & 78.8502$^{***}$ &  79.6360$^{***}$ &   75.3077$^{***}$ &    75.6079$^{***}$ &     74.6908$^{***}$ &      70.6229$^{***}$ &       82.5384$^{***}$\\        
                            &  (14.6076)  &   (14.7884)  &    (14.9298)  &     (14.9083)  &      (14.7132)  &       (14.9215)  &        (14.9656) \\         
Indirectly Covid-relevant    &             &              &               &    -38.5379    &                 &                  &       18.9492    \\        
                            &             &              &               &     (29.0879)  &                 &                  &        (30.7139) \\         
    \hline
    \hline
 Mainsection FE            &               &               &  O               &               &               &  O             &  O            \\
    \hline
N                         &  645          & 645           & 645              & 643           & 641           & 641           & 640          \\
df                        &  643          & 642           & 634              & 639           & 637           & 629           & 627          \\
RMSE                      &  0.3342       & 0.3338        & 0.3311           & 0.3282        & 0.3273        & 0.3249        & 0.3096       \\
Mean Abs. Deviation       &  0.1823       & 0.1823        & 0.1783           & 0.1792        & 0.1785        & 0.1746        & 0.1696       \\
Median Abs. Deviation     &  0.1079       & 0.1070        & 0.1008           & 0.1073        & 0.1055        & 0.0995        & 0.1017       \\
Error IQR                 &  0.2142       & 0.2140        & 0.2027           & 0.2150        & 0.2102        & 0.1985        & 0.2031       \\
   \hline
  \multicolumn{8}{l}{Signif. codes:  $^{***}$ $p<0.001$, $^{**}$ $p<0.01$, $^{*}$ $p<0.05$, $^{\dotr}$ $p<0.1$}\\
\end{tabular}
\caption{Robust regression with Huber's T norm with growth 2019 (in the year before the pandemic, i.e. 04/2019-03/2020) as PastGrowth.}
\label{tab:rob:g19}
\end{table}

\begin{figure}[H]
    \begin{center}
  \includegraphics[width=.7\textwidth]{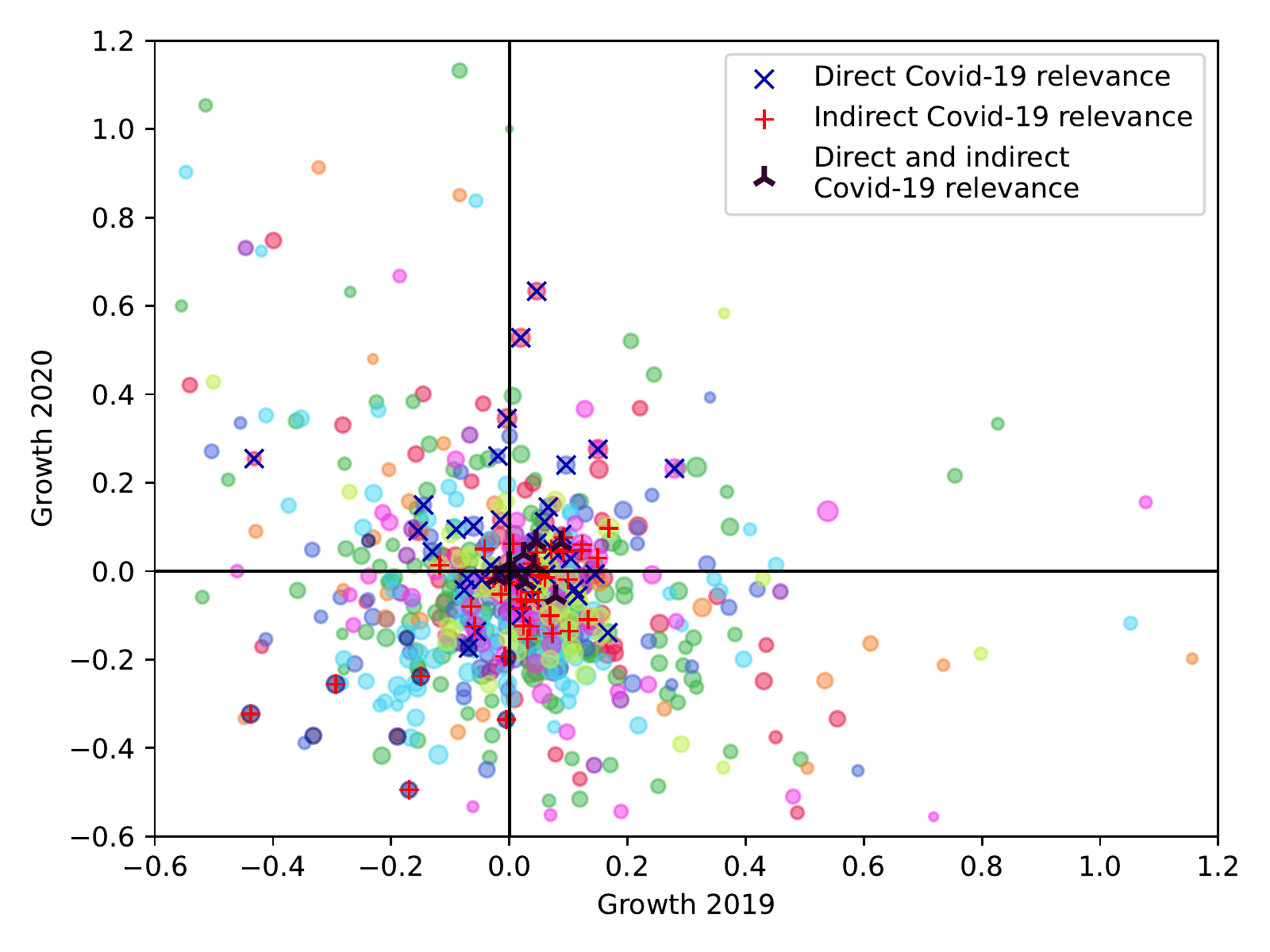}
\caption{Growth 2019 vs. growth 2020 by subclass with the size of the dot indicating the (weighted) number of patent applications per class (logarithmic scale). Different colors indicate different mainsections (see also figure~\ref{fig:growthplot:mainsections}).  Only the center of the distribution is shown. The 50 most directly and 50 most indirectly Covid-19 relevant subclasses are marked.}
\label{fig:growthplot}  
    \end{center}
\end{figure}

Indirect Covid-relevance, on the other hand, is not a significant influence - different from what the OLS fit suggested. This difference in growth patterns between directly and indirectly Covid-19 relevant subclasses is also clearly visible in figure~\ref{fig:growthplot}

The size of the technological field - $\log(NumberPreCovid)$ - also lost its significance for all models except model (7). This suggests an interaction of this variable with both indirect Covid-19 relevance and mainsection fixed effects. $PastGrowth$ in its specification as growth in the year before the pandemic (April 2019 to March 2020) is highly significant, but with different signs: The effect is positive in models (5) and (6), but negative in model (7). This again suggests an interaction with both indirect Covid-19 relevance and mainsection fixed effects.

\begin{table}[H]
\footnotesize
\centering
\begin{tabular}{lccccccc}
  \hline
                          & (1)           & (2)           & (3)              & (4)           & (5)           & (6)           & (7)          \\ 
  \hline
Mainsection[T.B]             &             &              &   -0.0454$^{\dotr}$   &                &                 &      -0.0435$^{\dotr}$   &       -0.0430$^{\dotr}$  \\        
                            &             &              &    (0.0247)   &                &                 &       (0.0244)   &        (0.0244)  \\         
Mainsection[T.C]             &             &              &   -0.0399     &                &                 &      -0.0385     &       -0.0495$^{\dotr}$  \\        
                            &             &              &    (0.0281)   &                &                 &       (0.0277)   &        (0.0278)  \\         
Mainsection[T.D]             &             &              &   -0.0499     &                &                 &      -0.0400     &       -0.0188    \\        
                            &             &              &    (0.0368)   &                &                 &       (0.0366)   &        (0.0368)  \\         
Mainsection[T.E]             &             &              &   -0.0163     &                &                 &      -0.0166     &       -0.0183    \\        
                            &             &              &    (0.0383)   &                &                 &       (0.0377)   &        (0.0378)  \\         
Mainsection[T.F]             &             &              &   -0.0696$^{*}$  &                &                 &      -0.0702$^{**}$ &       -0.0730$^{**}$\\        
                            &             &              &    (0.0272)   &                &                 &       (0.0268)   &        (0.0269)  \\         
Mainsection[T.G]             &             &              &   -0.0111     &                &                 &      -0.0204     &       -0.0165    \\        
                            &             &              &    (0.0283)   &                &                 &       (0.0283)   &        (0.0285)  \\         
Mainsection[T.H]             &             &              &   -0.0084     &                &                 &      -0.0109     &       -0.0021    \\        
                            &             &              &    (0.0329)   &                &                 &       (0.0325)   &        (0.0328)  \\         
Mainsection[T.Y]             &             &              &   -0.2694$^{***}$ &                &                 &      -0.2643$^{***}$ &       -0.3204$^{***}$\\        
                            &             &              &    (0.0580)   &                &                 &       (0.0572)   &        (0.0617)  \\         
   \hline
  \multicolumn{8}{l}{Signif. codes:  $^{***}$ $p<0.001$, $^{**}$ $p<0.01$, $^{*}$ $p<0.05$, $^{\dotr}$ $p<0.1$}\\
\end{tabular}
\caption{Mainsection fixed effect coeffficients in robust regressions with Huber's T norm with growth 2019 (in the year before the pandemic, i.e. 04/2019-03/2020) as PastGrowth.}
\label{tab:rob:g19:msFE}
\end{table}

\begin{figure}[H]
\begin{center}
\centering
 \subfloat[A - human necessities]{\includegraphics[width=.32\textwidth]{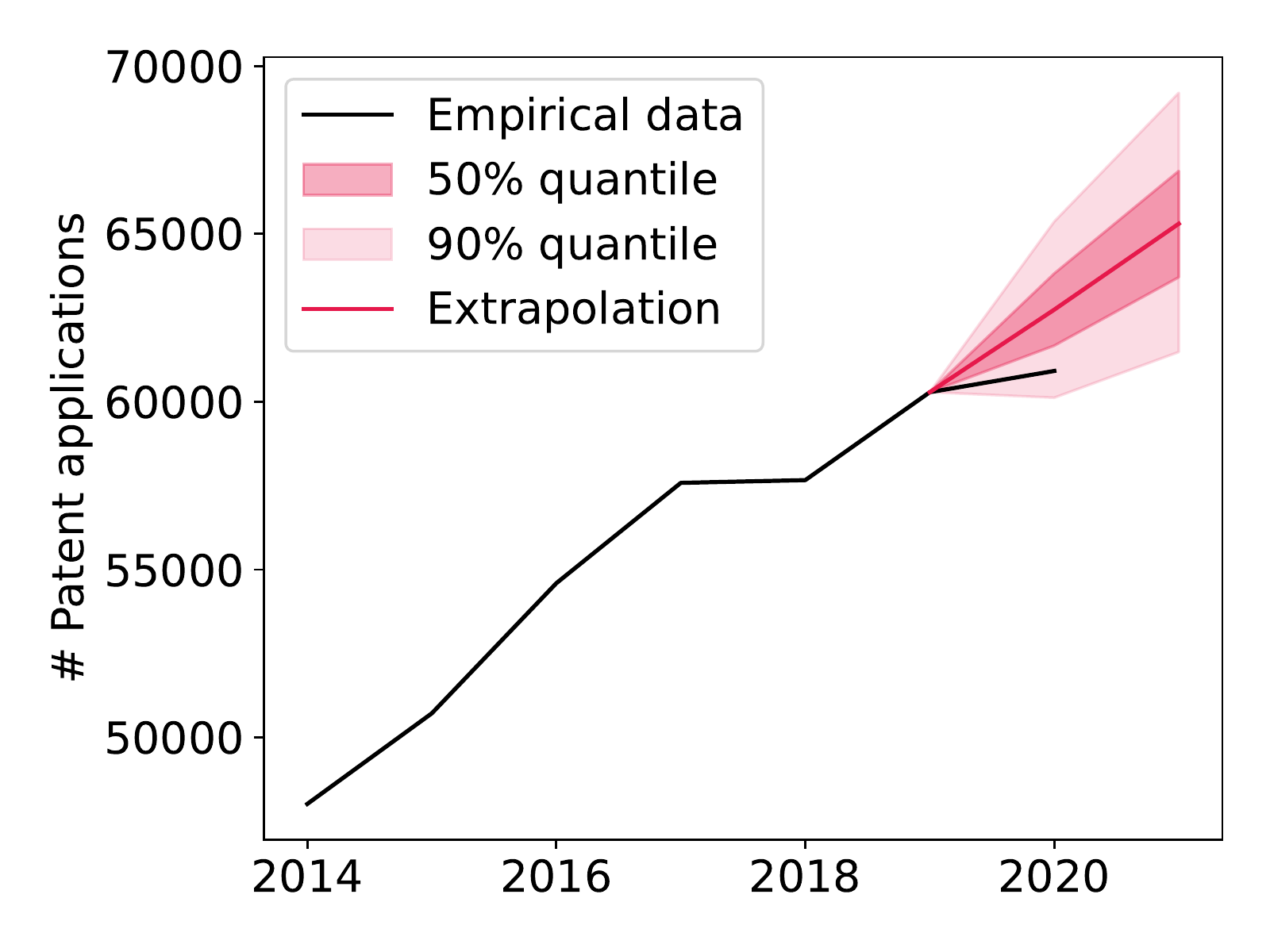}}
 \subfloat[B - operations, transport]{\includegraphics[width=.32\textwidth]{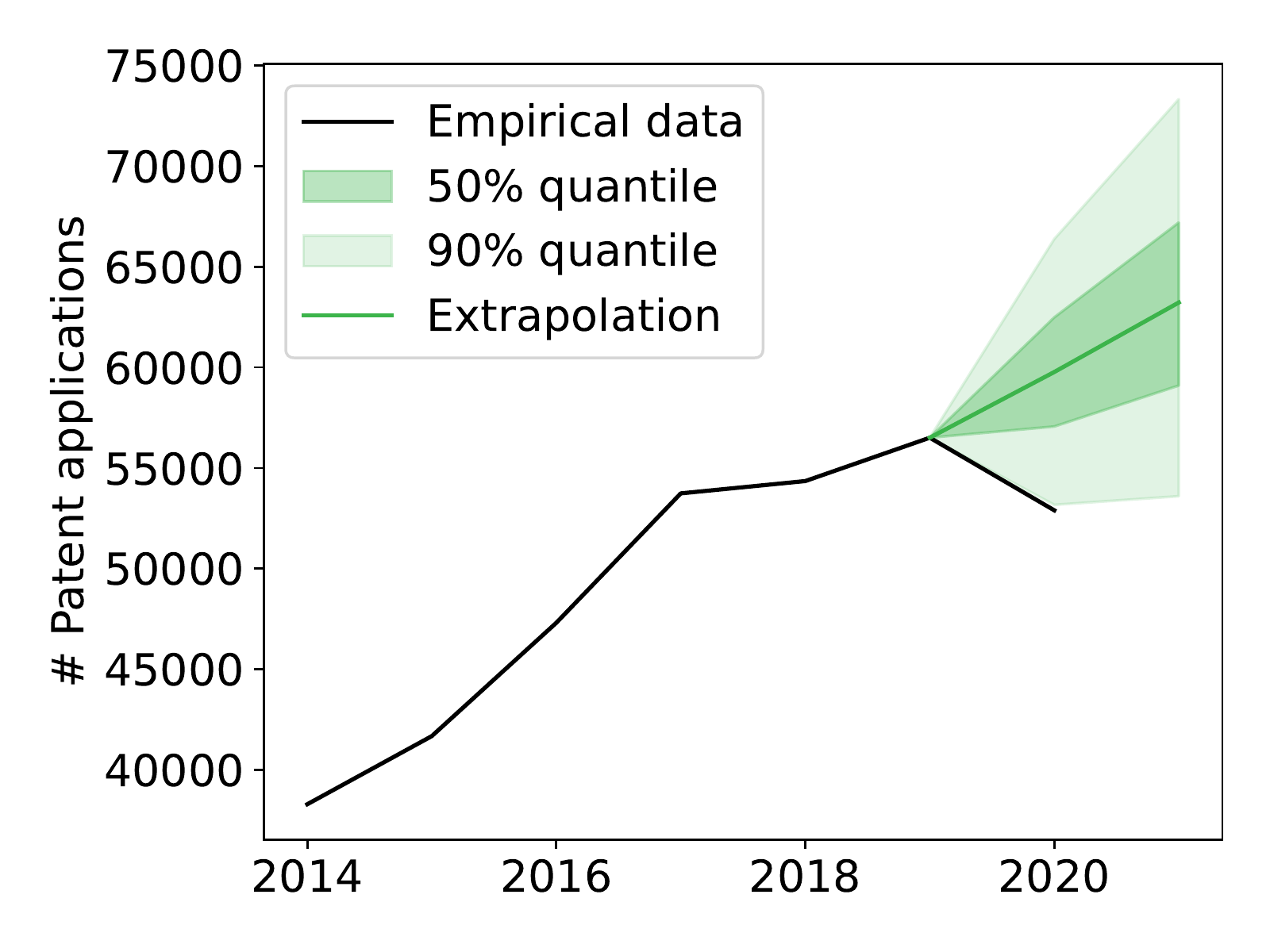}}
 \subfloat[C - chemistry, metallurgy]{\includegraphics[width=.32\textwidth]{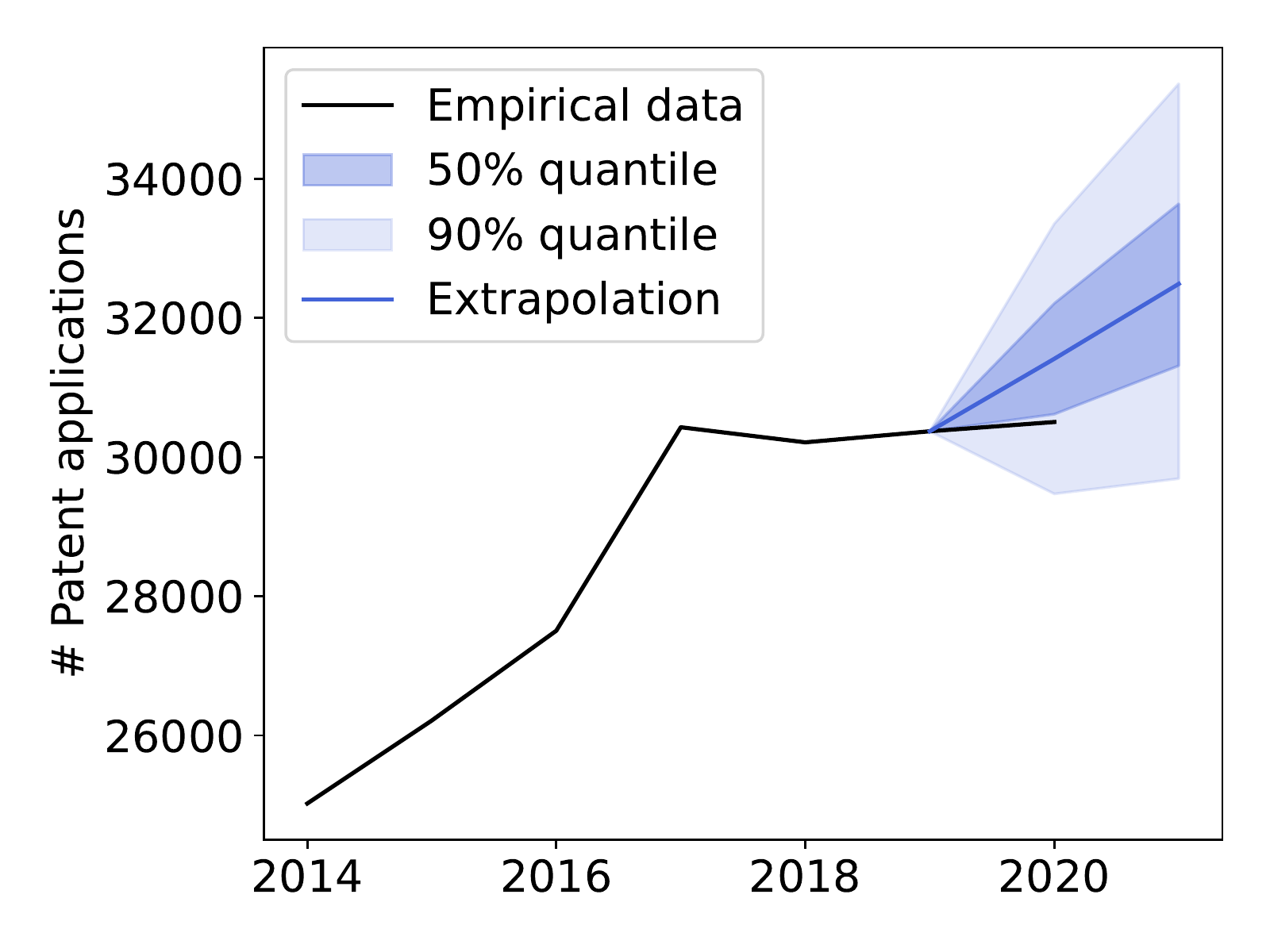}}\\
 \subfloat[D - textiles, paper]{\includegraphics[width=.32\textwidth]{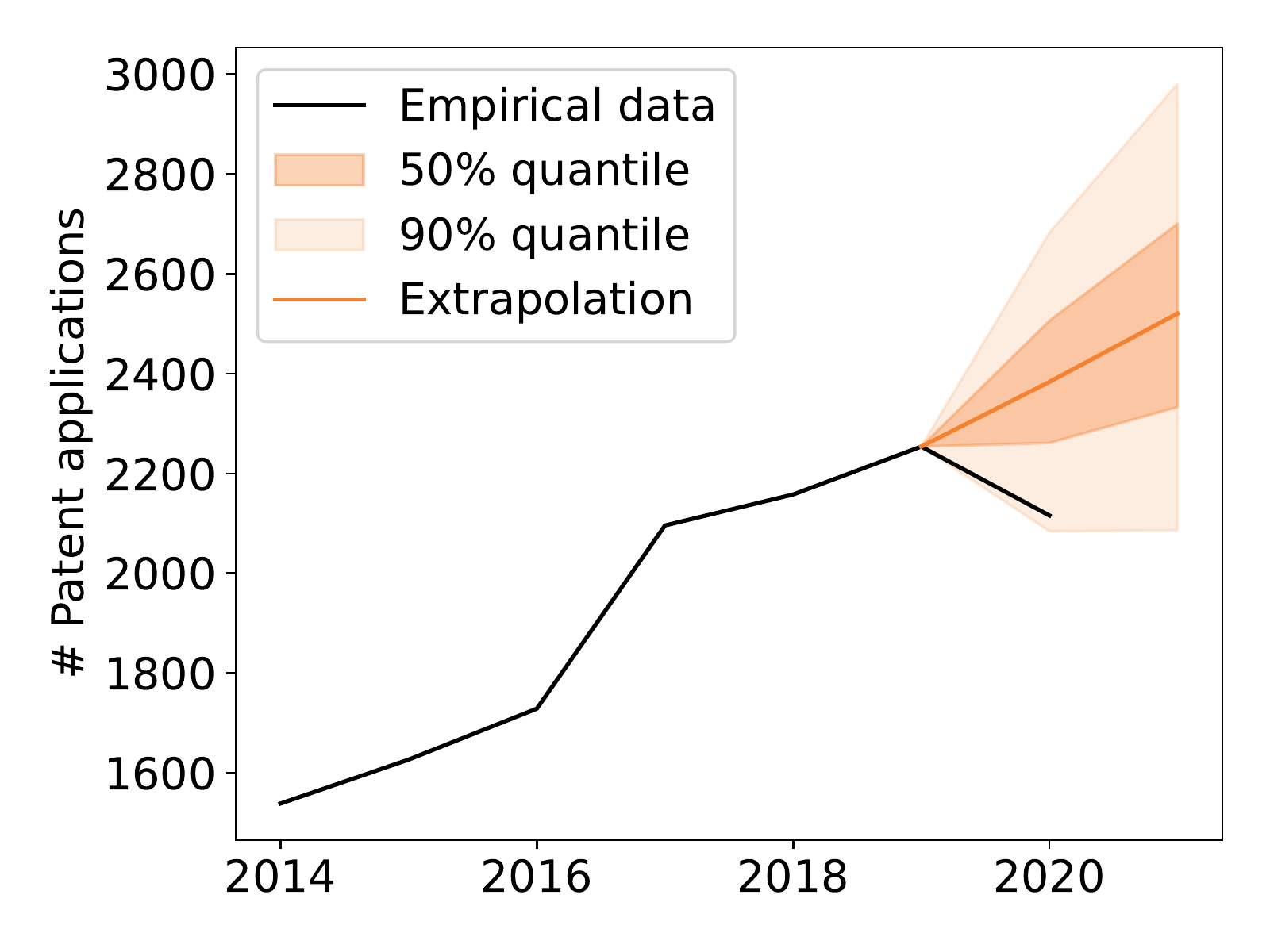}}
 \subfloat[E - fixed constructions]{\includegraphics[width=.32\textwidth]{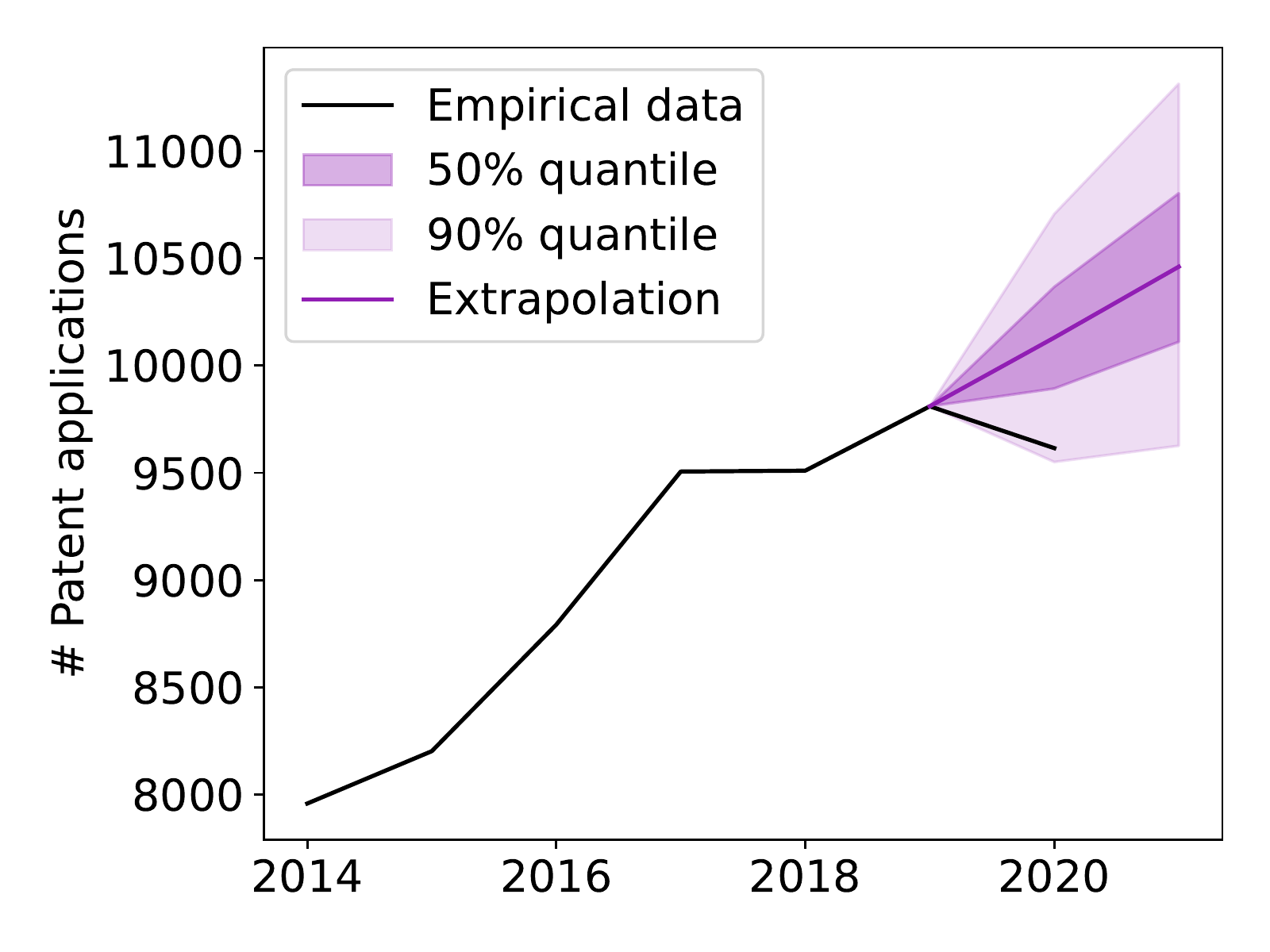}}
 \subfloat[F - engineering]{\includegraphics[width=.32\textwidth]{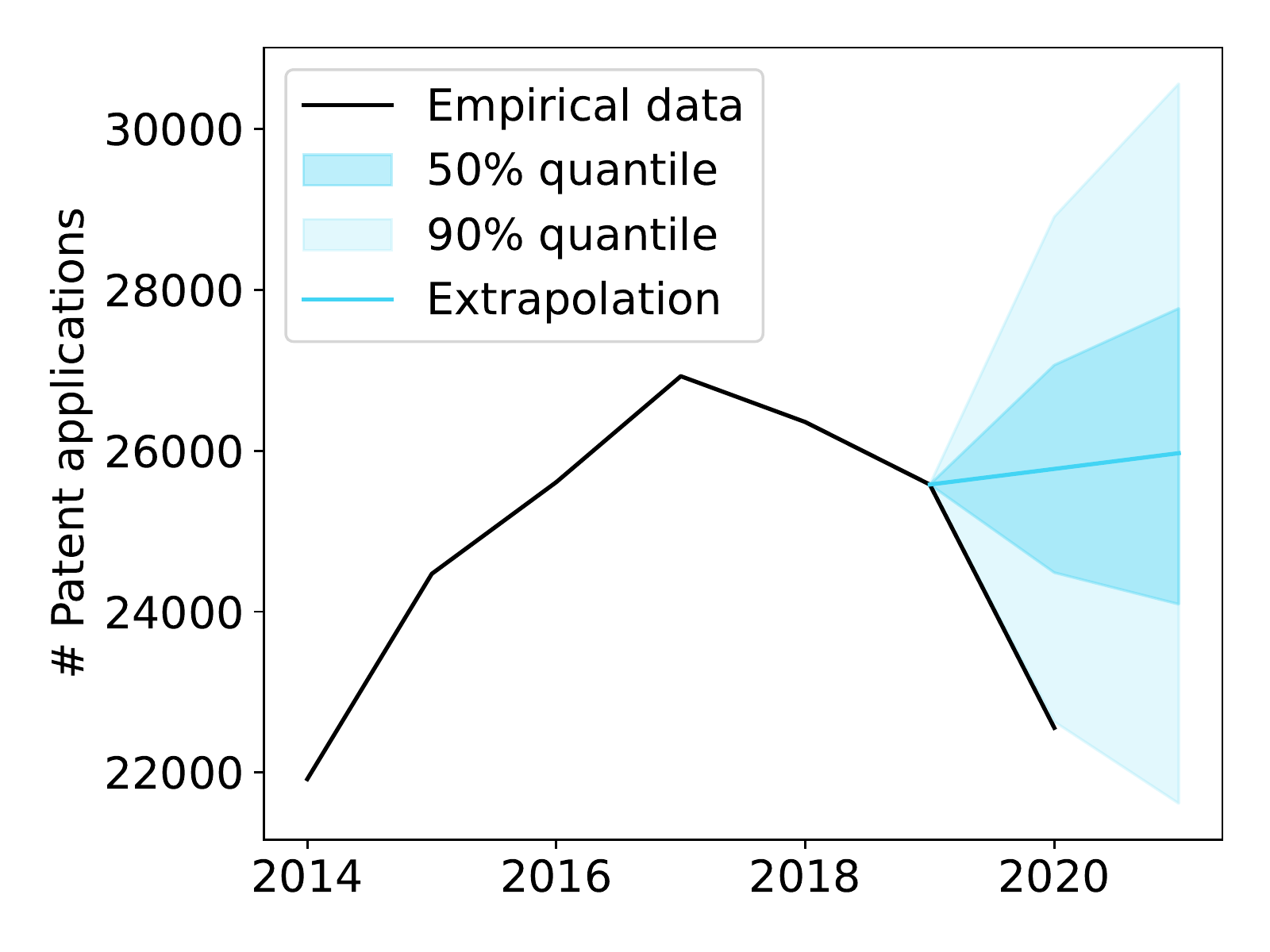}}\\
 \subfloat[G - physics]{\includegraphics[width=.32\textwidth]{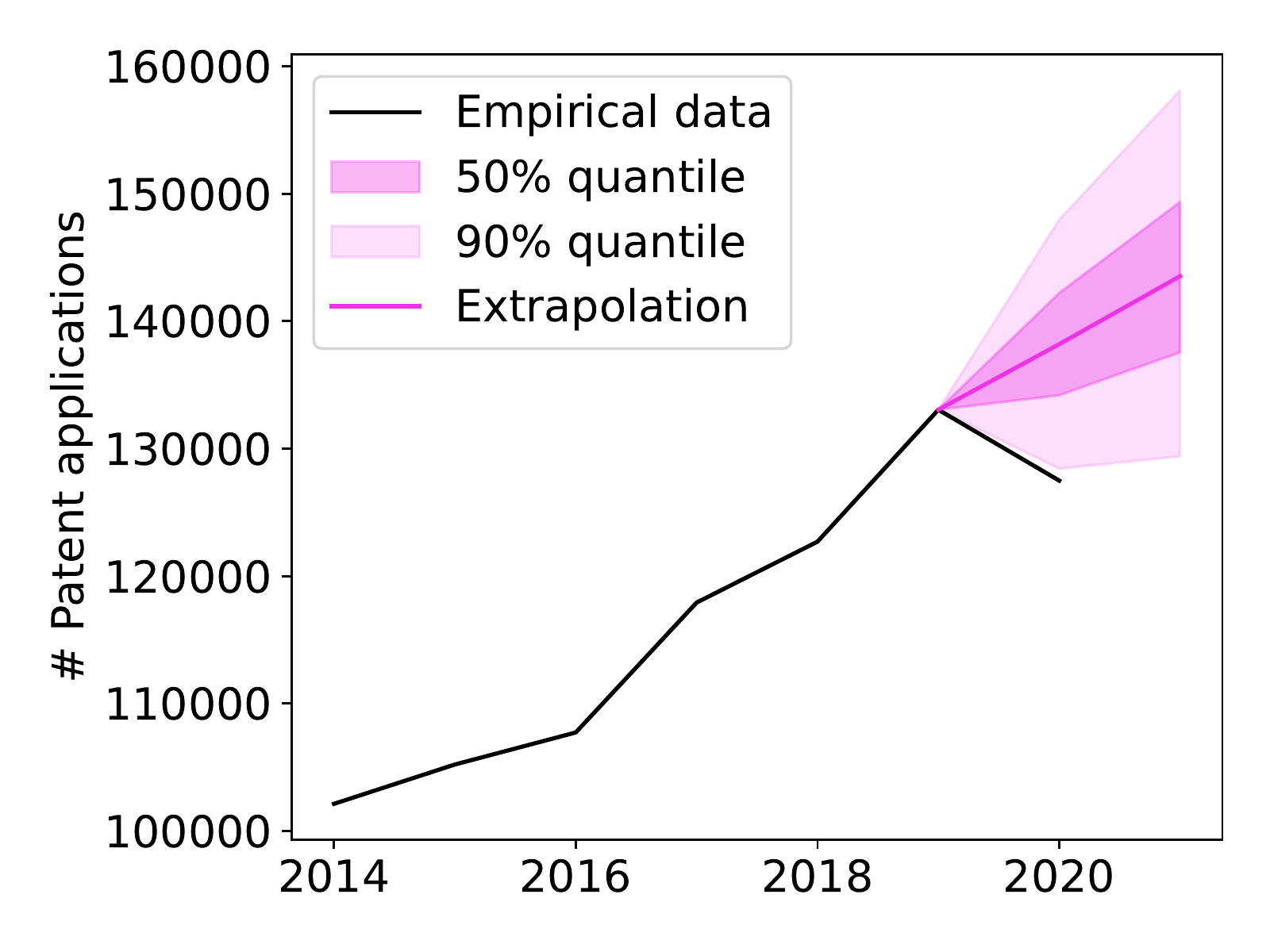}}
 \subfloat[H - electricity]{\includegraphics[width=.32\textwidth]{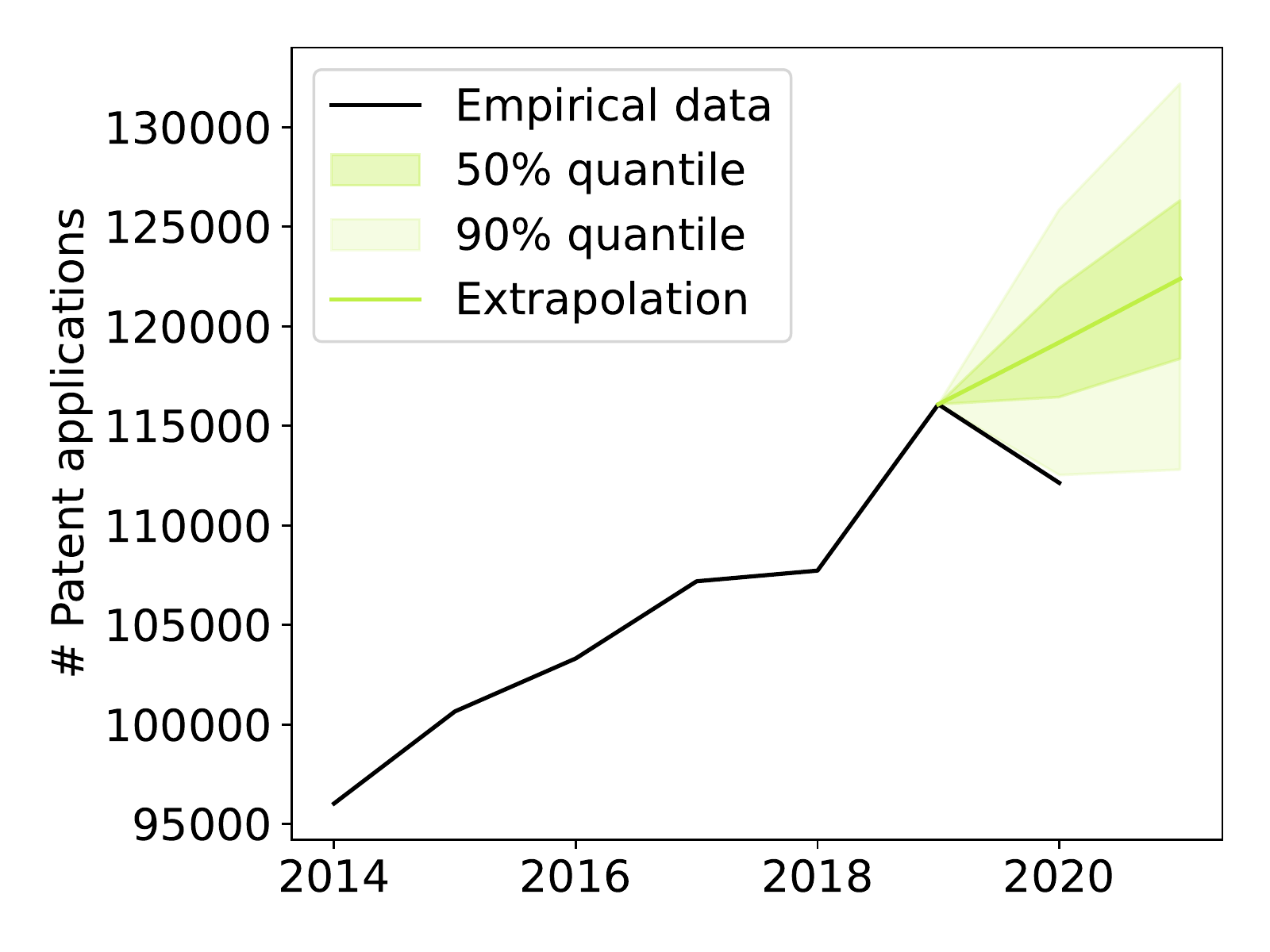}}
 \subfloat[Y - new technology tags]{\includegraphics[width=.32\textwidth]{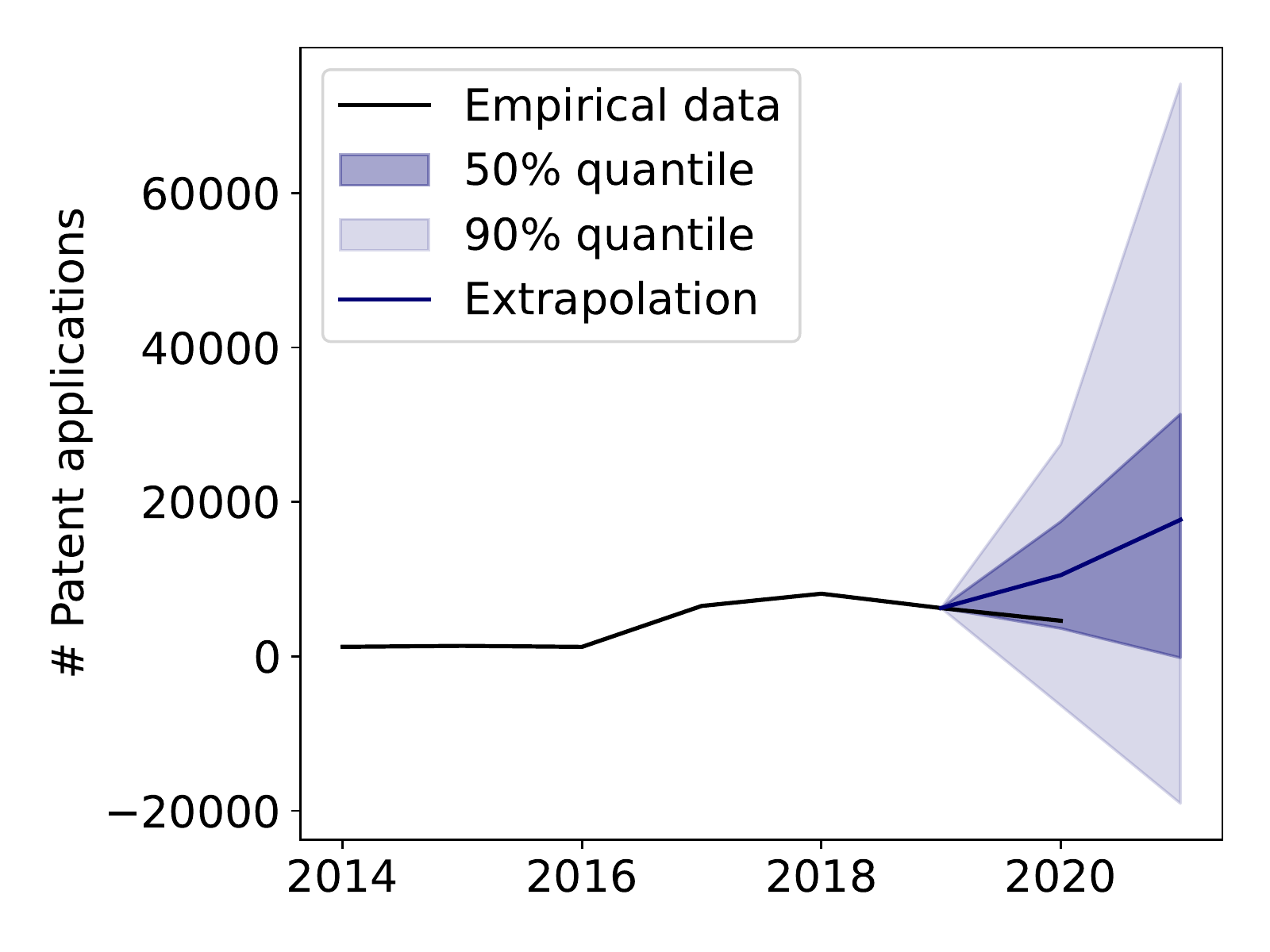}}
\caption{Patent applications by year and mainsection, extrapolation and realization for 2020 and 2021.}
\label{fig:growth:mainsections}  
\end{center}
\end{figure}

\begin{figure}[H]
\centering
 \subfloat[A - human necessities]{\includegraphics[width=.32\textwidth]{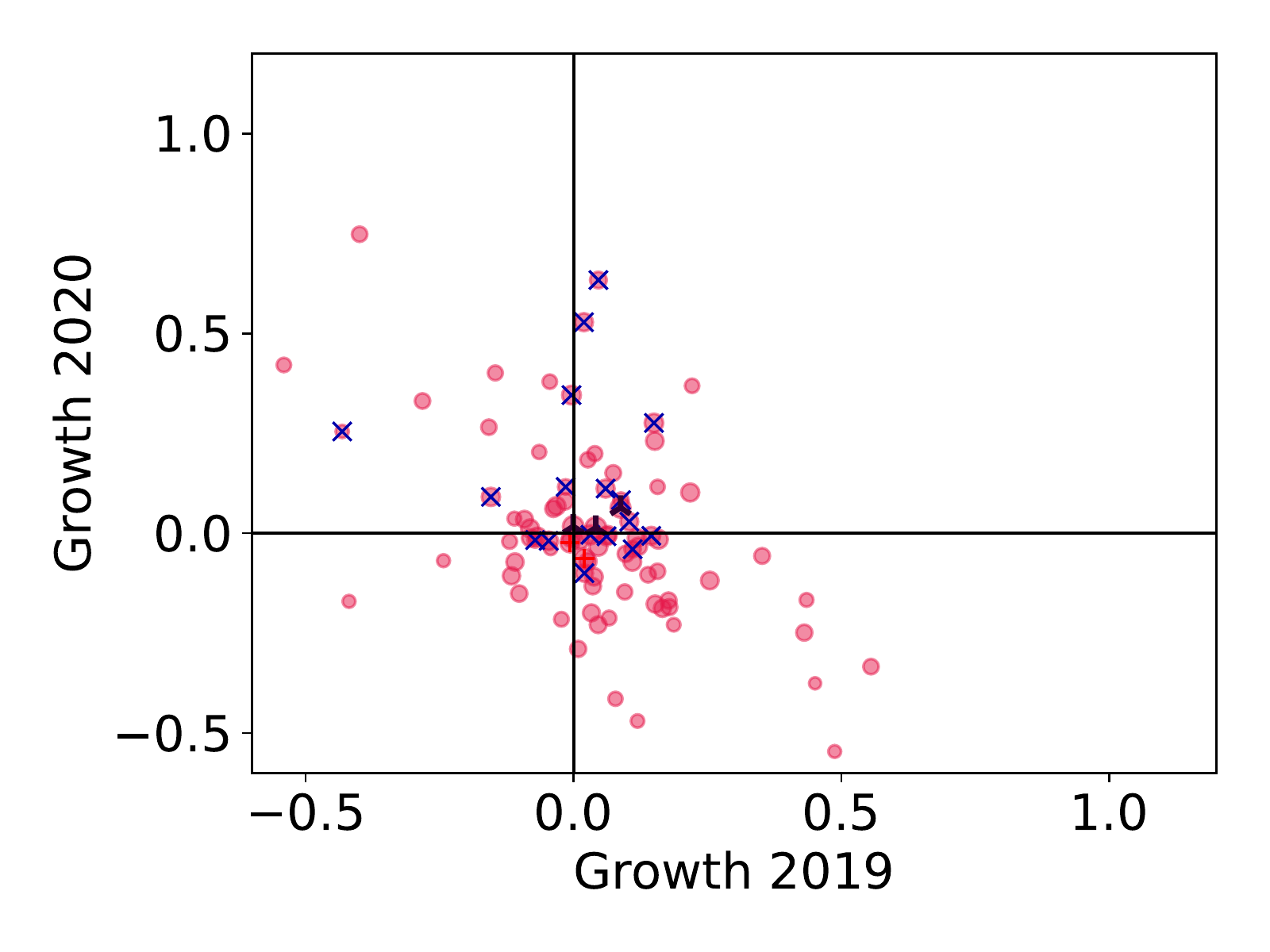}}
 \subfloat[B - operations, transport]{\includegraphics[width=.32\textwidth]{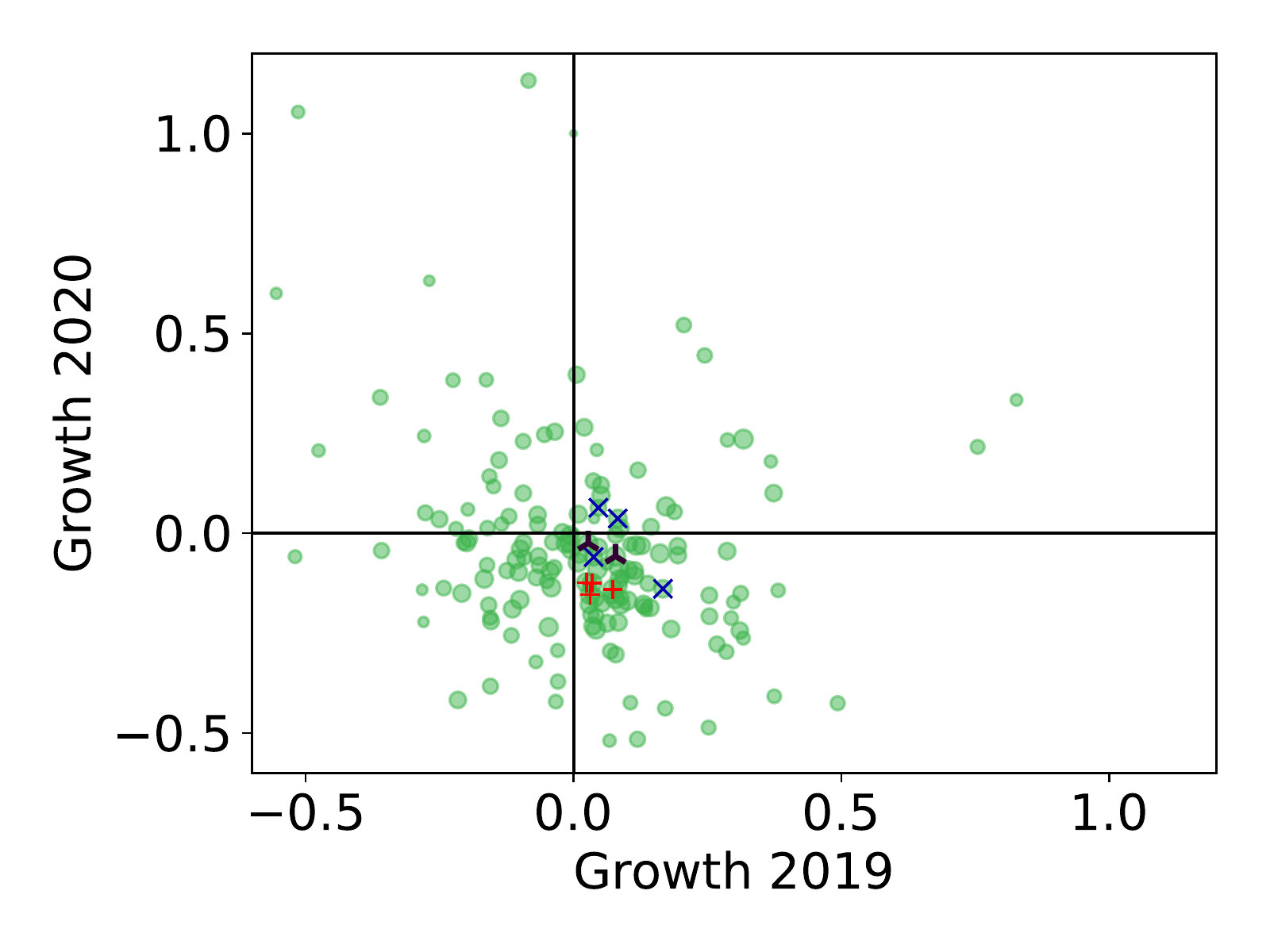}}
 \subfloat[C - chemistry, metallurgy]{\includegraphics[width=.32\textwidth]{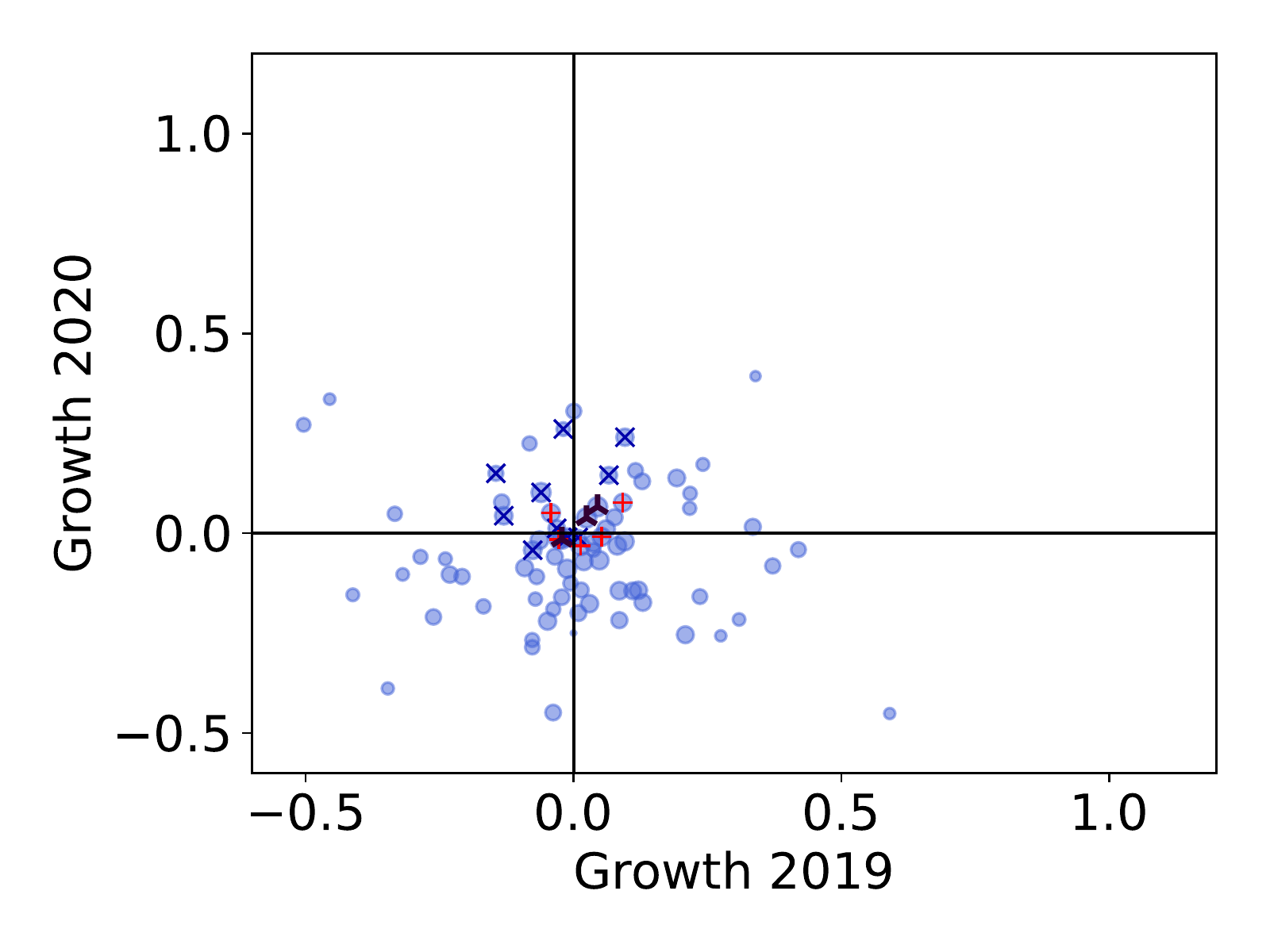}}
\\
 \subfloat[D - textiles, paper]{\includegraphics[width=.32\textwidth]{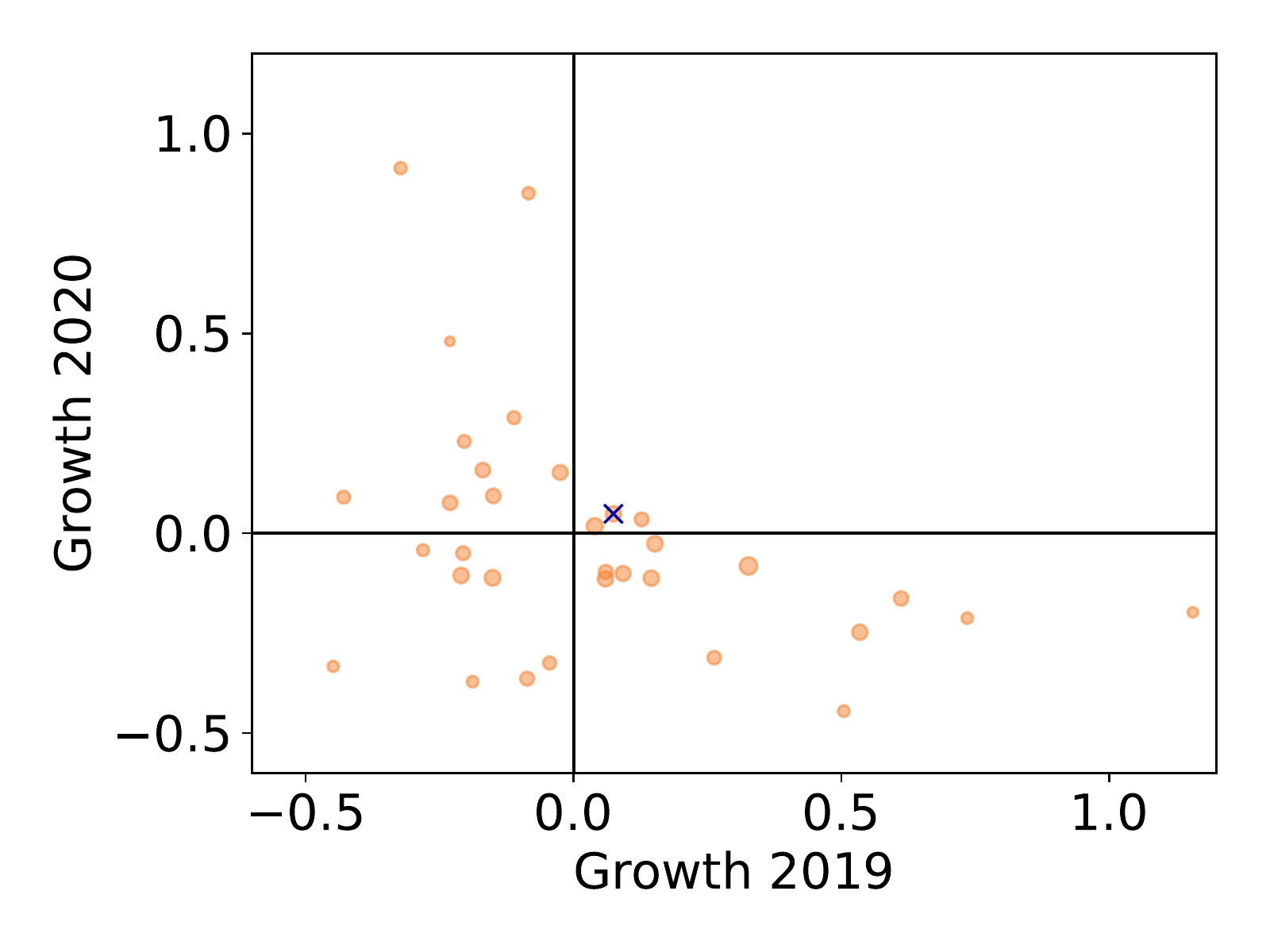}}
 \subfloat[E - fixed constructions]{\includegraphics[width=.32\textwidth]{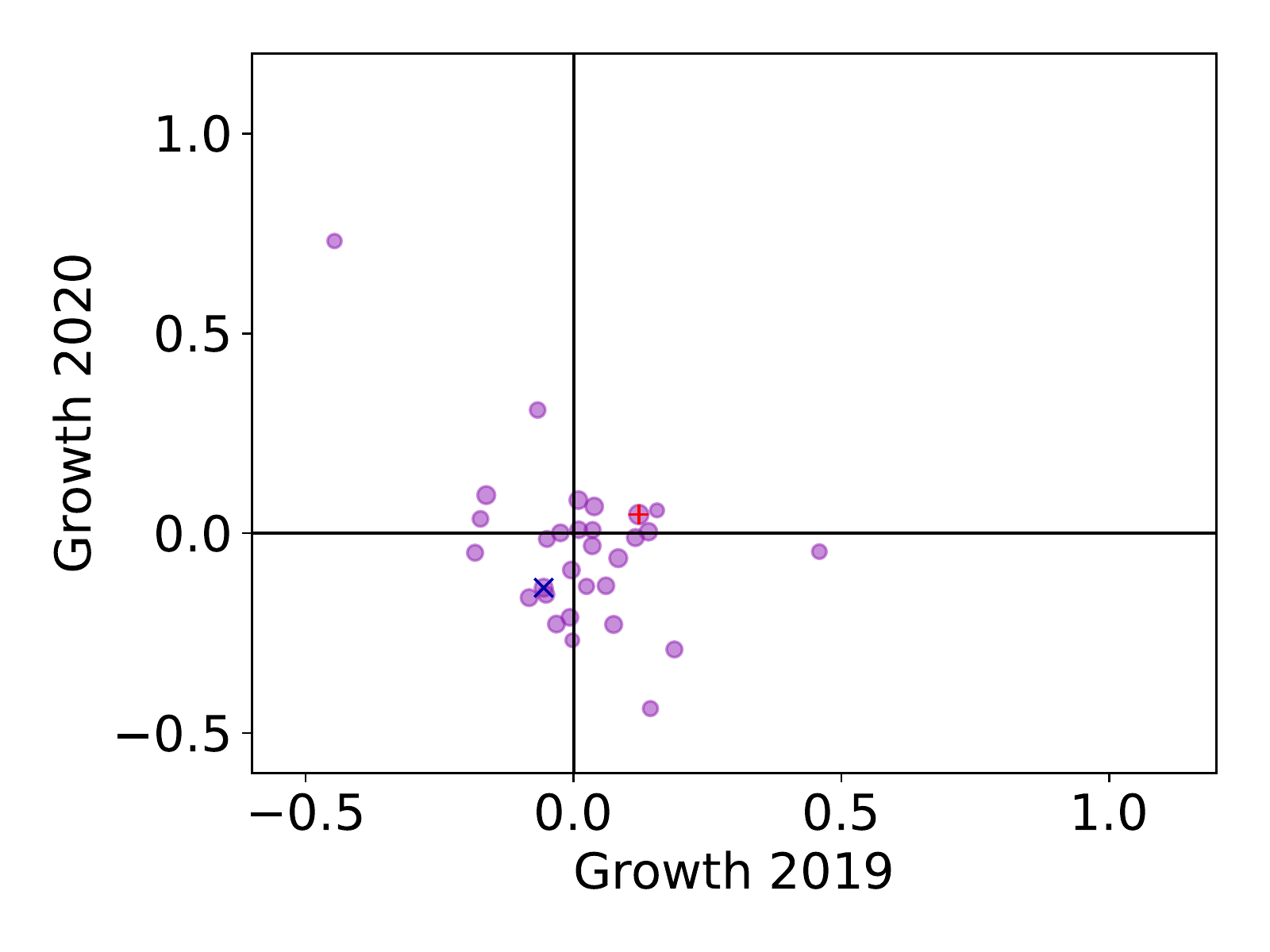}}
 \subfloat[F - engineering]{\includegraphics[width=.32\textwidth]{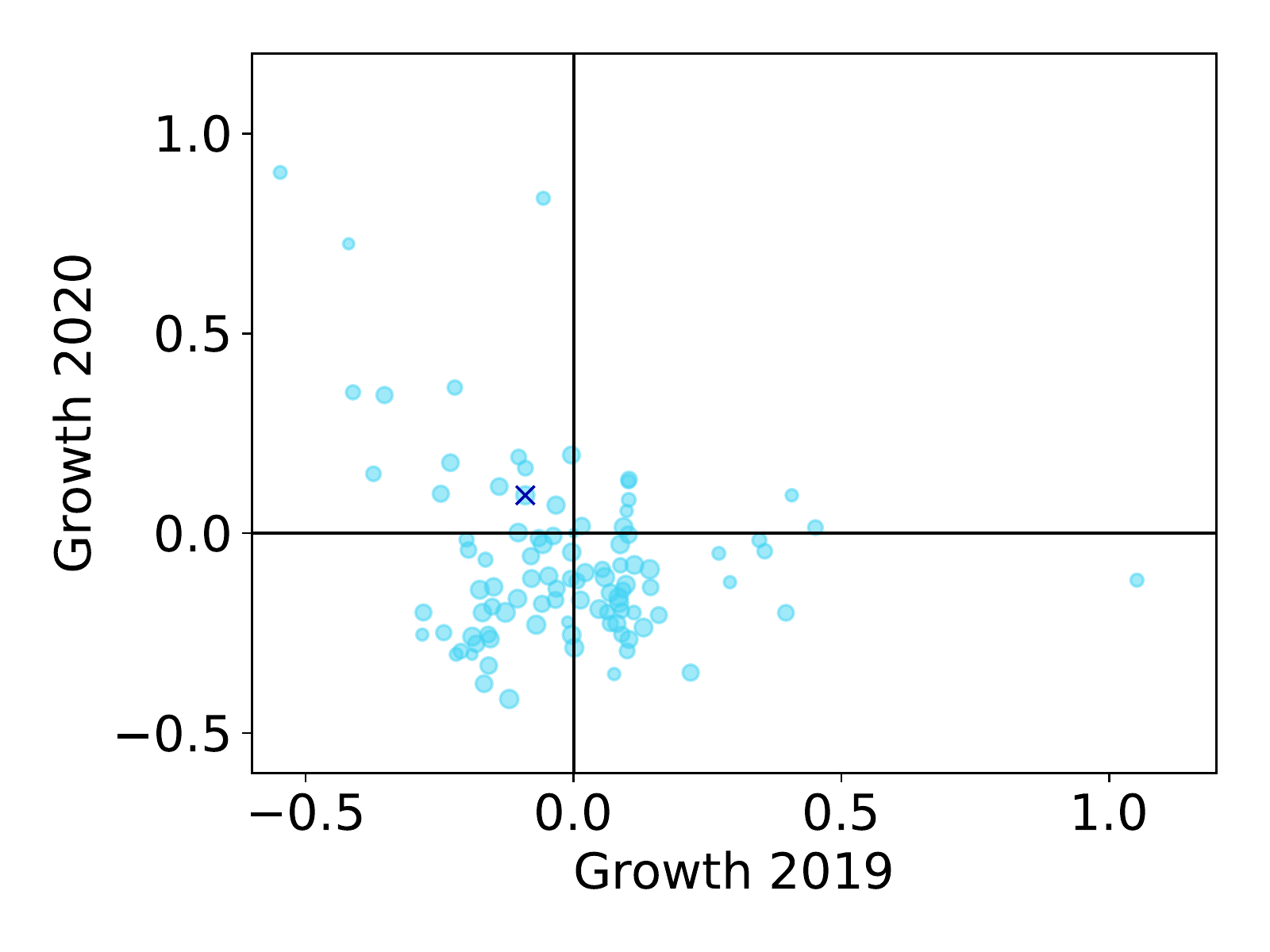}}
\\
 \subfloat[G - physics]{\includegraphics[width=.32\textwidth]{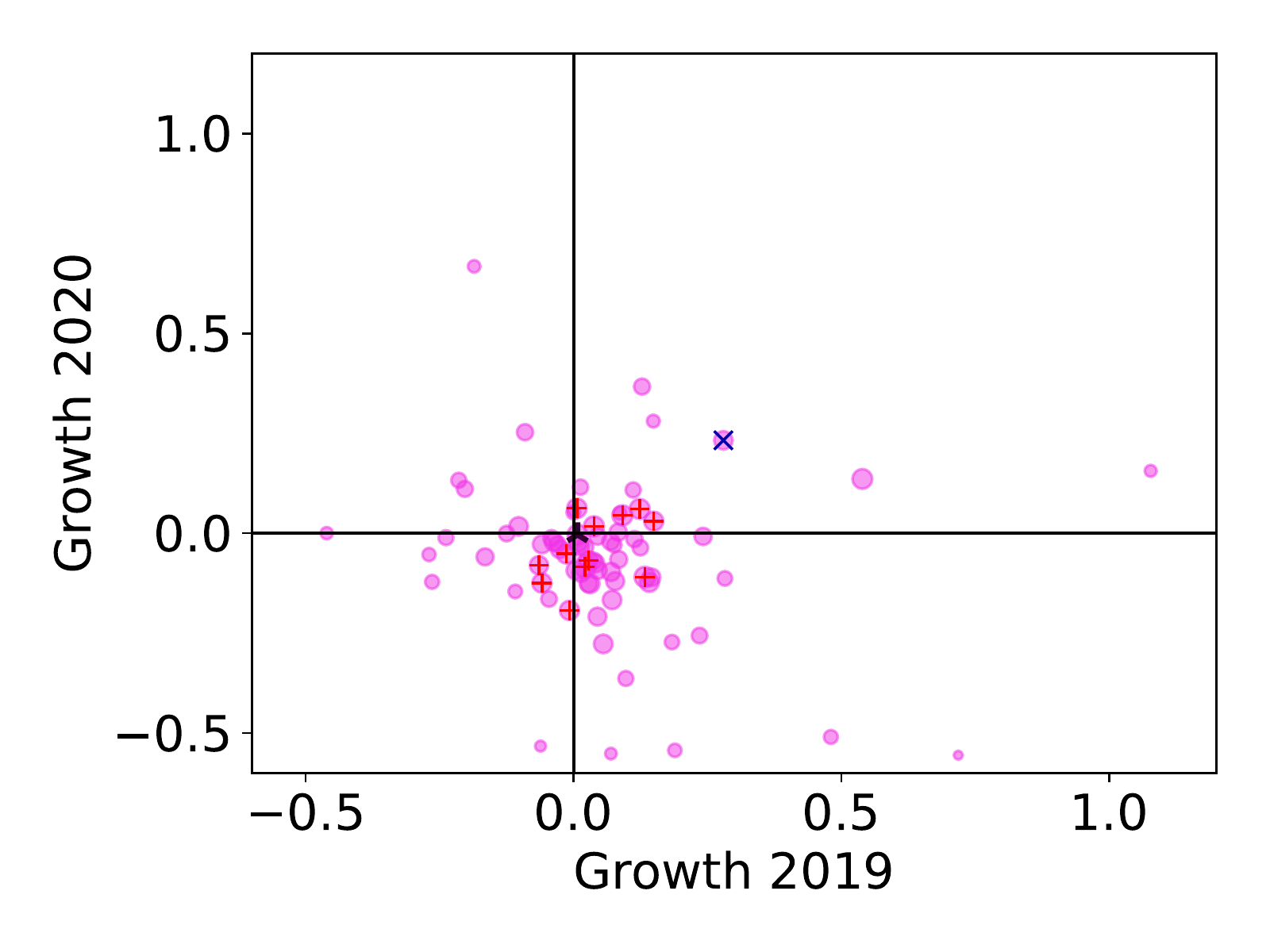}}
 \subfloat[H - electricity]{\includegraphics[width=.32\textwidth]{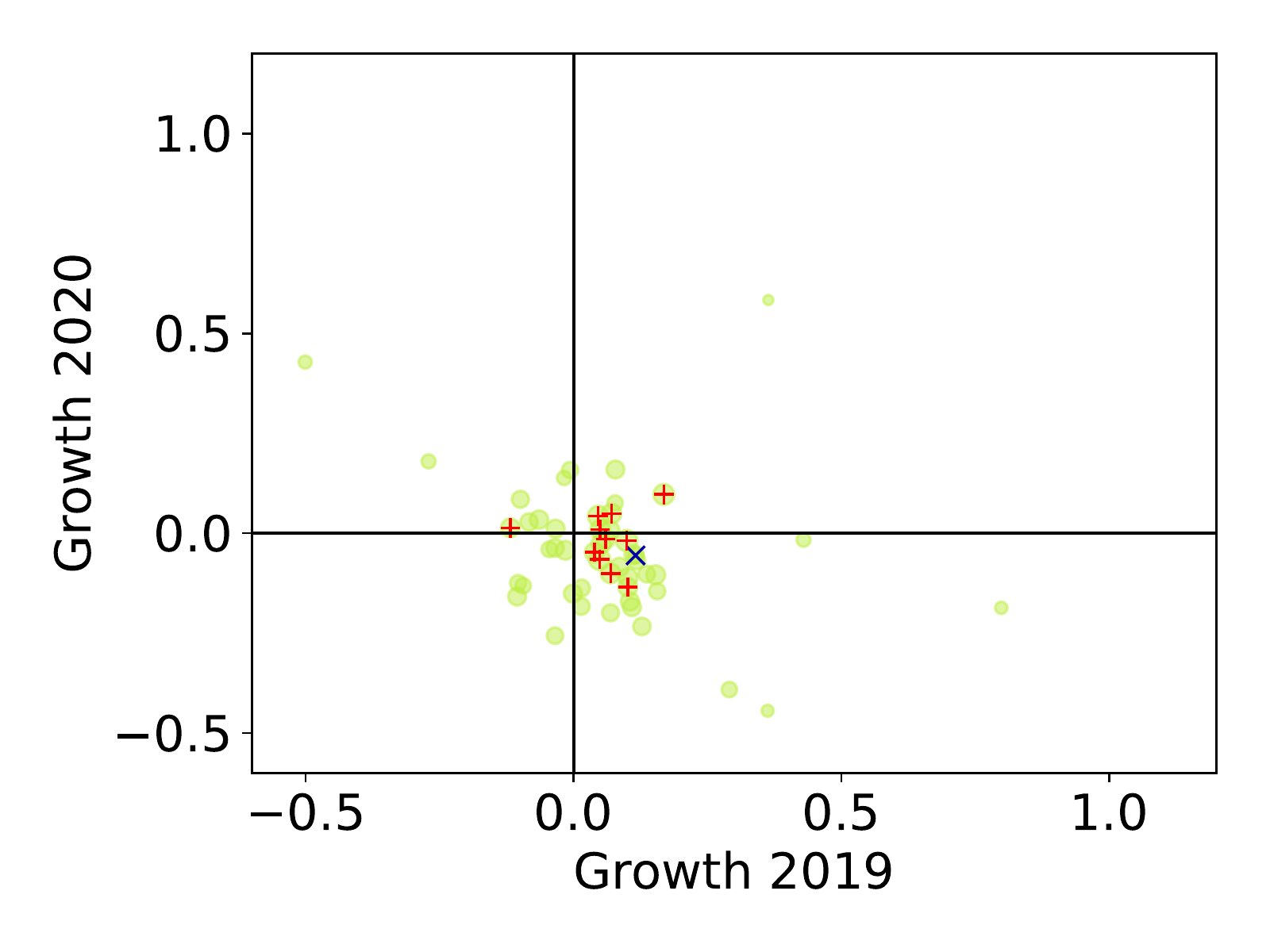}}
 \subfloat[Y - new technology tags]{\includegraphics[width=.32\textwidth]{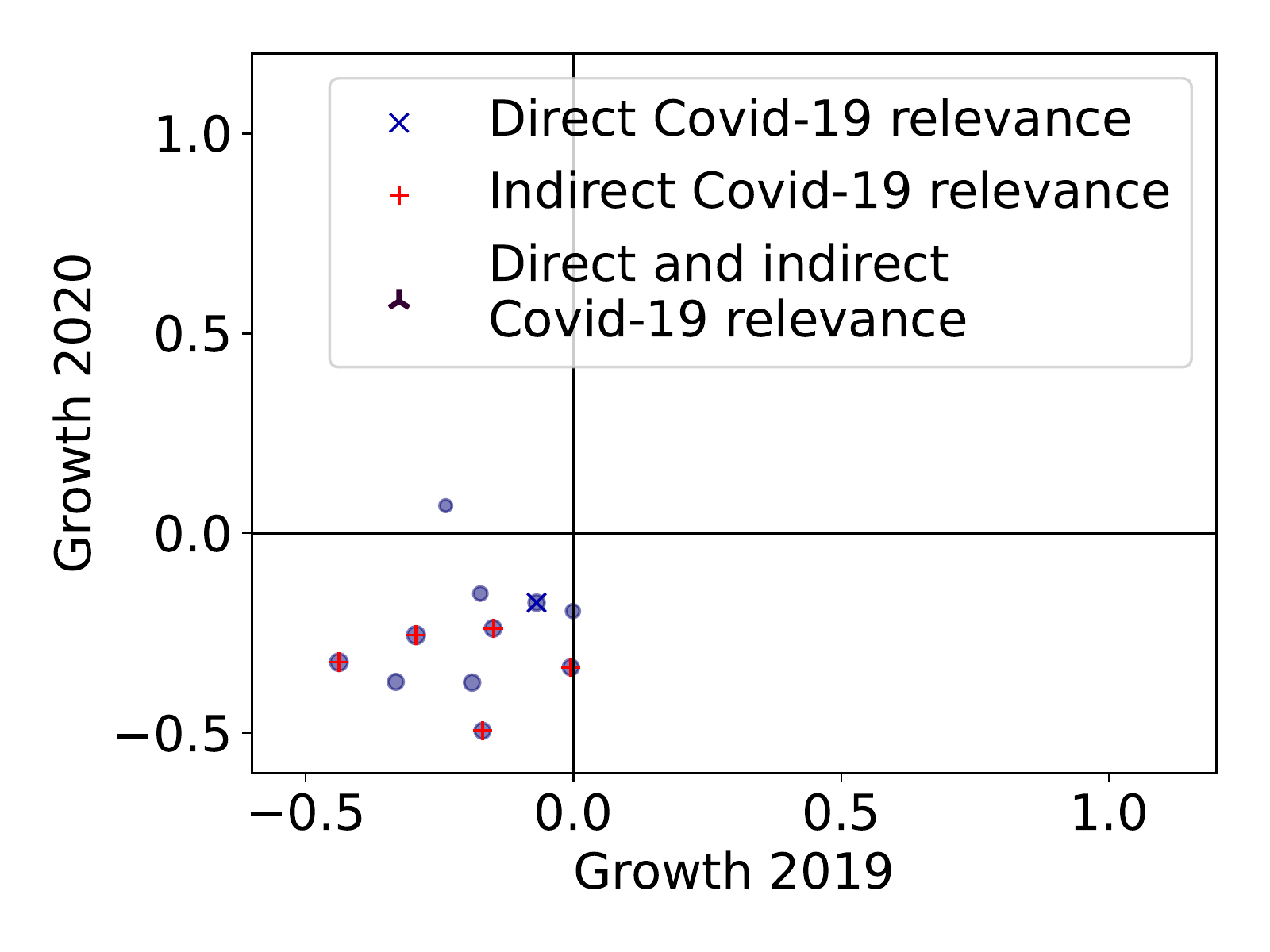}}
\caption{Growth 2019 vs. growth 2020 by subclass and mainsection (except Y) with the size of the dot indicating the (weighted) number of patent applications per class (logarithmic scale). Only the center of the distribution is shown. The 50 most directly and 50 most indirectly Covid-19 relevant subclasses are marked.}
\label{fig:growthplot:mainsections}  
\end{figure}

 CPC mainsection\footnote{We provide the list of the mainsections in table~\ref{tab:mainsections}; for details on the CPC classification system, see \citet{USPTO/EPO17}.} fixed effects (see table~\ref{tab:rob:g19:msFE}) show significant differences between mainsection A and B, A and F, A and Y, as well as, for model (7) only, A and C. Since mainsection A, human necessities, is one of the fields of research that did continue to grow into the pandemic (see figure~\ref{fig:growth:mainsections}), it is no surprise that all these effects are negative (i.e., B, F, Y, and perhaps C perform less well than A). The marginal significance (at the $10\%$ level) of the mainsection C fixed effect in model (7) is likely another interaction with indirect Covid-19 relevance as well as the PastGrowth term and is thus to be taken with a grain of salt. The differences in performance between the mainsections is also visible in figures~\ref{fig:growth:mainsections} and \ref{fig:growthplot:mainsections}.

We find that additional regressors tend to improve all four goodness variables we consider (RMSE, mean and median absolute deviation, and error IQR) except in case of model (7). Median absolute deviation and error IQR increase again in model (7). In fact, according to these two measures, model (7) performs worse than model (3), indicating that there is a problem here and the sign change of $PastGrowth$ (in the specification of growth in 2019) should also be treated with scepticism.

\subsection{Robustness check}
\label{sect:robustnesscheck}

\paragraph{A binary indicator for Covid-19 relevance }

\begin{table}[H]
\footnotesize
    \begin{center}
\begin{tabular}{lccccccc}
  \hline
                          & (1)           & (2)           & (3)              & (4)           & (5)           & (6)             & (7)  \\ 
  \hline
(Intercept)                  & -0.0762$^{***}$ &  -0.0698$^{**}$ &   -0.0153     &    -0.0679$^{**}$ &     -0.0728$^{**}$ &      -0.0209     &       -0.0088    \\        
                            &  (0.0075)   &   (0.0238)   &    (0.0320)   &     (0.0256)   &      (0.0242)   &       (0.0322)   &        (0.0343)  \\         
log(Number pre-Covid)        &             &  -0.0010     &   -0.0049     &    -0.0013     &     -0.0005     &      -0.0038     &       -0.0058    \\        
                            &             &   (0.0037)   &    (0.0039)   &     (0.0042)   &      (0.0038)   &       (0.0040)   &        (0.0044)  \\         
Growth 2019                  &             &              &               &                &     0.0155$^{***}$  &      0.0152$^{***}$  &       0.0152$^{***}$ \\        
                            &             &              &               &                &      (0.0020)   &       (0.0020)   &        (0.0020)  \\         
Covid-relevant               & 0.1396$^{***}$  &  0.1406$^{***}$  &   0.1353$^{***}$  &    0.1401$^{***}$  &     0.1308$^{***}$  &      0.1247$^{***}$  &       0.1234$^{***}$ \\        
                            &  (0.0271)   &   (0.0275)   &    (0.0279)   &     (0.0276)   &      (0.0277)   &       (0.0284)   &        (0.0286)  \\         
Indirectly Covid-relevant    &             &              &               &    0.0049      &                 &                  &       0.0237     \\        
                            &             &              &               &     (0.0309)   &                 &                  &        (0.0312)  \\         
   \hline
Mainsection FE            &               &               &  O               &                &               &  O               &  O              \\
   \hline
N                         &  645          & 645           & 645              & 645            & 641           & 641             & 641 \\
df                        &  643          & 642           & 634              & 641            & 637           & 629             & 628 \\
RMSE                      &  0.3339       & 0.3335        & 0.3311           & 0.3335        & 0.3276        & 0.3256        & 0.3251       \\
Mean Abs. Deviation       &  0.1823       & 0.1824        & 0.1785           & 0.1823        & 0.1786        & 0.1750        & 0.1749       \\
Median Abs. Deviation     &  0.1069       & 0.1080        & 0.1002           & 0.1063        & 0.1052        & 0.0998        & 0.0982       \\
Error IQR                 &  0.2110       & 0.2128        & 0.2013           & 0.2130        & 0.2040        & 0.1986        & 0.1948       \\
   \hline
  \multicolumn{7}{l}{Signif. codes:  $^{***}$ $p<0.001$, $^{**}$ $p<0.01$, $^{*}$ $p<0.05$, $^{\dotr}$ $p<0.1$}\\
\end{tabular}
\caption{Robust regression with Huber's T norm with growth 2019 (in the year before the pandemic, i.e. 04/2019-03/2020) as PastGrowth and binary indicators for Covid-19 relevance.}
\label{tab:rob:incid:g19}
\end{center}
\end{table}

While continuous Covid-19 relevance measures (regression table~\ref{tab:rob:g19}) carry more detailed information and should therefore generally be preferred, we do consider regressions with binary Cobid-19 measures (regression table~\ref{tab:rob:incid:g19}) as a point of comparison, since these allow for a more illustrative interpretation. The intercept in table~\ref{tab:rob:incid:g19} is negative where it is significant (models (1), (2), (4), (5)), even in the simplest model without other effects (model (1)), indicating a generally negative growth in 2020. As Covid-19 relevance is binary, the sum of the two coefficients yields the growth estimate for directly Covid-19 relevant fields of technology. Since the coefficient for Covid-19 relevance is positive and, indeed, positive by roughly double the amount that of the intercept is negative, the model predicts solid growth in patenting for directly Covid-19 relevant fields. The signs and magnitudes of the two coefficients remain about the same for models (1), (2), (4), and (5). It should also be noted that the other remaining significant coefficients, that for $PastGrowth$ in the specification as growth in 2019 in models (5), (6), and (7), is relatively small compared to the intercept and the coefficient of Covid-19 relevance. In order to offset these larger effects, the growth rate in 2019 would have to reach several hundred percent. 

\paragraph{PastGrowth as 5 year average}

\begin{table}[H]
\footnotesize
\begin{center}
\begin{tabular}{lccccccc}
  \hline
                          & (1)           & (2)           & (3)              & (4)           & (5)           & (6)           & (7)          \\ 
  \hline
(Intercept)                  & -0.0734$^{***}$ &  -0.0648$^{**}$ &   -0.0022     &    -0.0916$^{***}$ &     -0.0649$^{**}$ &      -0.0057     &       -0.0037    \\        
                            &  (0.0075)   &   (0.0239)   &    (0.0323)   &     (0.0257)   &      (0.0244)   &       (0.0324)   &        (0.0345)  \\         
log(Number pre-Covid)        &             &  -0.0013     &   -0.0052     &    0.0034      &     -0.0013     &      -0.0045     &       -0.0051    \\        
                            &             &   (0.0037)   &    (0.0039)   &     (0.0043)   &      (0.0038)   &       (0.0040)   &        (0.0045)  \\         
Average growth 2014-2019     &             &              &               &                &     0.0000      &      0.0003      &       -0.0003    \\        
                            &             &              &               &                &      (0.0009)   &       (0.0009)   &        (0.0010)  \\         
Covid-relevant               & 78.8502$^{***}$ &  79.6360$^{***}$ &   75.3077$^{***}$ &    75.6079$^{***}$ &     75.9746$^{***}$ &      69.2170$^{***}$ &       74.3049$^{***}$\\        
                            &  (14.6076)  &   (14.7884)  &    (14.9298)  &     (14.9083)  &      (16.2776)  &       (16.5956)  &        (16.8415) \\         
Indirectly Covid-relevant    &             &              &               &    -38.5379    &                 &                  &       8.6341     \\        
                            &             &              &               &     (29.0879)  &                 &                  &        (30.7899) \\         
   \hline
Mainsection FE            &               &               &  O               &               &               &  O             &  O            \\
   \hline
N                         &  645          & 645           & 645              & 643           & 641           & 641           & 640          \\
df                        &  643          & 642           & 634              & 639           & 637           & 629           & 627          \\
RMSE                      &  0.3342       & 0.3338        & 0.3311           & 0.3282        & 0.3313        & 0.3286        & 0.3230       \\
Mean Abs. Deviation       &  0.1823       & 0.1823        & 0.1783           & 0.1792        & 0.1798        & 0.1757        & 0.1736       \\
Median Abs. Deviation     &  0.1079       & 0.1070        & 0.1008           & 0.1073        & 0.1049        & 0.0993        & 0.0990       \\
Error IQR                 &  0.2142       & 0.2140        & 0.2027           & 0.2150        & 0.2100        & 0.1988        & 0.1977       \\
   \hline
  \multicolumn{8}{l}{Signif. codes:  $^{***}$ $p<0.001$, $^{**}$ $p<0.01$, $^{*}$ $p<0.05$, $^{\dotr}$ $p<0.1$}\\
\end{tabular}
\caption{Robust regression with Huber's T norm with average growth 2015-2019 as PastGrowth.}
\label{tab:rob:avg}
\end{center}
\end{table}

For another robustness check, we use the average annual growth in the 5 years leading up to the pandemic for the $PastGrowth$ variable. The results mostly stay the same except that the average annual growth loses significance.

\subsection{Discussion}

Our analysis shows in line with what scholars expected \citep{Gaudry/Gianola20} that growth in patenting took a substantial hit during the pandemic. As shown in figure~\ref{fig:growth:mainsections}, this is directly evident in all mainsections except A (human necessities) and C (chemistry, metallurgy). Of course, these two are the fields, in which vaccines, drugs, chemicals used for testing disinfectants (mainsection C) as well as the techniques to deploy these in medicine (mainsection A) would be registered, so their relatively strong performance might, in fact, be related to Covid-19 relevance. We were able to show that this holds definitely at the subclass level: directly Covid-19 relevant subclasses performed significantly better in terms of growth in the number of patents in the first year of the pandemic. The effect was found to be strong enough to raise the baseline expectations of growth to positive levels for directly Covid-19 relevant subclasses. This does, however, not extend to indirectly Covid-19 relevant subclasses, for which there is no significant effect. 

The effect of past growth is significant and positive (likely an indicator of autocorrelation), but not large compared to the dominant effects of direct Covid-19 relevance and mainsection fixed effects.

It is debatable whether continuous and strong development of medical and other Covid-19 relevant technology aided the later response to the pandemic. While there is a strong Pearson correlation between $CovidRelevant$ and $PastGrowth$ in the specification as average growth 2015-2019 (correlation coefficient $\varrho=0.42$), this correlation disappears in the more robust Spearman rank correlation (coefficient $r=-0.07$).

\section{Conclusion}
\label{sect:conclusion}

Efforts to quickly and redundantly deploy applications of earlier groundwork in response to the Covid-19 pandemic would have been primary objectives of innovation policy in crisis mode \citep{Gross/Sampat21}. Such efforts are evident in the difference between the effects of direct and indirect Covid-19 relevance found above, and they seem to have been successful. 

In this paper, we provided measures for direct and indirect relevance of technological fields to managing the Covid-19 pandemic at the CPC subclass level. While we found no effect for indirect Covid-19 relevance, direct Covid-19 relevance has a strong, positive, and highly significant effect. The order of magnitude of this effect, measured in terms of the percentage point increase in annual growth of patent application numbers per CPC subclass, is on par with the share of patent applications in the subclass that do refer to Covid-19 explicitly.

We found an additional fixed effect of technological fields - in this case the CPC mainsection. Mainsections A (human necessities) and C (chemistry, metallurgy), for instance, had a stronger performance than others like B (operations, transport) or F (engineering). This could indicate spillovers in fields that are intuitively relevant to, e.g., vaccine development (mainsection C) and deployment (mainsection A). It could also indicate that our keyword search does not capture all innovations relevant to pandemic management (this is evident) and introduces a slight bias (which is less evident).

Of course, it is known that there were efforts to develop technology to guide and aid the response to Covid-19 - vaccines in particular - as is analyzed in \citet{Florio22}. However, it is by no means obvious that this would be visible and significant in patent application numbers. Should the generated knowledge indeed be published - a requirement for most patent applications after 18 months - and thus shared with the competition? Do the developers see it as possible and expedient to gainfully implement their technology themselves to license it to others? The answers are, evidently, yes.

Our analysis suggests that efforts to re-focus the innovation system on quickly providing applied solutions to urgent problems of the management of the pandemic were successful. At the same time, other fields have taken a hit, including more basic research that is - indirectly - relevant to the management of the pandemic. The continuing threat Covid-19 and potentially other pandemics pose emphasizes that it is important that research and innovation in these indirectly pandemic-relevant fields regains the strength it had before 2020. Our analysis does not permit us to give an estimate, to what extent innovation related to the management of Covid-19 was publicly funded and should therefore be in the public domain, as other authors \cite{Florio22} suggest. However, the instruments for this - requiring placing patents from publicly funded research in the public domain - would be available for future efforts of short-term innovation guidance, if desired.

The question remains - what are the mid- and long-term effects? Will basic research, including indirectly Covid-19 relevant subsections, resume its former pace? Are current developments shaping the vaccination (and remote work?) technologies for decades to come, similar to the research conducted in World War II \citep{Lafondetal20}? Time - and careful observation of patent applications as they become public - will tell.

\section*{Acknowledgements}

Moritz Wilke has contributed to the publicly available code on \url{https://github.com/x0range/Covid19PatentApplications}, that has been used for the analysis. We are grateful for many helpful comments to Kerstin H\"otte, Sujung Jee, Mariia Shkolnykova, and Iciar Dominguez Lacasa as well as other participants of the EAEPE conference 2022. All remaining errors are our own.

\bibliographystyle{apalike}
\bibliography{main}

\begin{thebibliography}{}

\bibitem[Agarwal and Gaule, 2022]{agarwal2022drives}
Agarwal, R. and Gaule, P. (2022).
\newblock What drives innovation? lessons from covid-19 r\&d.
\newblock {\em Journal of Health Economics}, 82:102591.

\bibitem[Alexopoulos et~al., 2021]{alexopoulos2021evaluating}
Alexopoulos, M., Lyons, K., Mahetaji, K., and Chiu, K. (2021).
\newblock Evaluating the disruption of covid-19 on ai innovation using patent
  filings.
\newblock In {\em 2021 IEEE International Symposium on Technology and Society
  (ISTAS)}, pages 1--6. IEEE.

\bibitem[Amore, 2015]{amore2015companies}
Amore, M.~D. (2015).
\newblock Companies learning to innovate in recessions.
\newblock {\em Research Policy}, 44(8):1574--1583.

\bibitem[Archibugi et~al., 2013]{archibugi2013economic}
Archibugi, D., Filippetti, A., and Frenz, M. (2013).
\newblock Economic crisis and innovation: is destruction prevailing over
  accumulation?
\newblock {\em research Policy}, 42(2):303--314.

\bibitem[Arrow, 1962]{Arrow62}
Arrow, K. (1962).
\newblock {\em Economic Welfare and the Allocation of Resources for Invention},
  pages 609--626.
\newblock Princeton University Press.

\bibitem[Bachmann and Frutos-Bencze, 2022]{bachmann2022r}
Bachmann, P. and Frutos-Bencze, D. (2022).
\newblock R\&d and innovation efforts during the covid-19 pandemic: The role of
  universities.
\newblock {\em Journal of Innovation \& Knowledge}, 7(4):100238.

\bibitem[Berkes et~al., 2020]{Berkesetal20}
Berkes, E., Deschenes, O., Gaetani, R., Lin, J., and Severen, C. (2020).
\newblock Lockdowns and innovation: Evidence from the 1918 flu pandemic.
\newblock Working Paper 28152, National Bureau of Economic Research.

\bibitem[Bloom et~al., 2021]{Bloometal21}
Bloom, N., Davis, S.~J., and Zhestkova, Y. (2021).
\newblock Covid-19 shifted patent applications toward technologies that support
  working from home.
\newblock {\em AEA Papers and Proceedings}, 111:263--66.

\bibitem[Brem et~al., 2020]{brem2020impact}
Brem, A., Nylund, P., and Viardot, E. (2020).
\newblock The impact of the 2008 financial crisis on innovation: A dominant
  design perspective.
\newblock {\em Journal of Business Research}, 110:360--369.

\bibitem[Brem et~al., 2021]{brem2021implications}
Brem, A., Viardot, E., and Nylund, P.~A. (2021).
\newblock Implications of the coronavirus (covid-19) outbreak for innovation:
  Which technologies will improve our lives?
\newblock {\em Technological forecasting and social change}, 163:120451.

\bibitem[Dahlke et~al., 2021]{dahlke2021crisis}
Dahlke, J., Bogner, K., Becker, M., Schlaile, M.~P., Pyka, A., and Ebersberger,
  B. (2021).
\newblock Crisis-driven innovation and fundamental human needs: A typological
  framework of rapid-response covid-19 innovations.
\newblock {\em Technological Forecasting and Social Change}, 169:120799.

\bibitem[Florio, 2021]{Florio21}
Florio, M. (2021).
\newblock Patents for covid-19 vaccines are based on public research: a case
  study on the privatization of knowledge.
\newblock {\em CIRIEC International, Universit\'{e} de Li\`{e}ge Working
  Paper}, (WP2021-03).

\bibitem[Florio, 2022]{Florio22}
Florio, M. (2022).
\newblock To what extent patents for covid-19 mrna vaccines are based on public
  research and taxpayers’ funding? a case study on the privatization of
  knowledge.
\newblock {\em Industrial and Corporate Change}, 31(5):1137--1151.

\bibitem[Gaudry and Gianola, 2020]{Gaudry/Gianola20}
Gaudry, K. and Gianola, A. (2020).
\newblock Technology-specific patent filing trends during the pandemic.
\newblock {\em ipwatchdog.com}.
\newblock
  https://www.ipwatchdog.com/2020/11/01/technology-specific-patent-filing-trends-pandemic/id=126901/.

\bibitem[Gross and Sampat, 2021]{Gross/Sampat21}
Gross, D.~P. and Sampat, B.~N. (2021).
\newblock The economics of crisis innovation policy: A historical perspective.
\newblock {\em AEA Papers and Proceedings}, 111:346--50.

\bibitem[Guderian et~al., 2021]{guderian2021innovation}
Guderian, C.~C., Bican, P.~M., Riar, F.~J., and Chattopadhyay, S. (2021).
\newblock Innovation management in crisis: patent analytics as a response to
  the covid-19 pandemic.
\newblock {\em R\&D Management}, 51(2):223--239.

\bibitem[Han and Qian, 2020]{Han/Qian20}
Han, H. and Qian, Y. (2020).
\newblock Did enterprises’ innovation ability increase during the covid-19
  pandemic? evidence from chinese listed companies.
\newblock {\em Asian Economics Letters}, 1(3).

\bibitem[Huber, 1973]{Huber73}
Huber, P.~J. (1973).
\newblock Robust regression: Asymptotics, conjectures and monte carlo.
\newblock {\em The Annals of Statistics}, 1(5):799--821.

\bibitem[Jarque and Bera, 1980]{jarque1980efficient}
Jarque, C.~M. and Bera, A.~K. (1980).
\newblock Efficient tests for normality, homoscedasticity and serial
  independence of regression residuals.
\newblock {\em Economics letters}, 6(3):255--259.

\bibitem[Kim et~al., 2016]{kim2016impact}
Kim, B., Kim, E., Miller, D.~J., and Mahoney, J.~T. (2016).
\newblock The impact of the timing of patents on innovation performance.
\newblock {\em Research Policy}, 45(4):914--928.

\bibitem[Lafond et~al., 2020]{Lafondetal20}
Lafond, F., Greenwald, D.~S., and Farmer, J.~D. (2020).
\newblock Can stimulating demand drive costs down? world war ii as a natural
  experiment.
\newblock {\em INET Working Paper 2020-02, University of Oxford}.

\bibitem[Nelson, 1959]{Nelson59}
Nelson, R.~R. (1959).
\newblock The simple economics of basic scientific research.
\newblock {\em Journal of Political Economy}, 67(3):297--306.

\bibitem[Sampat and Shadlen, 2021]{Sampat/Shadlen21}
Sampat, B.~N. and Shadlen, K.~C. (2021).
\newblock The covid-19 innovation system.
\newblock {\em Health Affairs}, 40(3):400--409.
\newblock PMID: 33539184.

\bibitem[Seabold and Perktold, 2010]{Seabold/Perktold10}
Seabold, S. and Perktold, J. (2010).
\newblock statsmodels: Econometric and statistical modeling with python.
\newblock In {\em 9th Python in Science Conference}.

\bibitem[{U.S. Patent and Trademark Office} and {European Patent Office},
  2017]{USPTO/EPO17}
{U.S. Patent and Trademark Office} and {European Patent Office} (2017).
\newblock Guide to the cpc (cooperative patent classification), revision 4.00.
\newblock Available online,
  https://www.cooperativepatentclassification.org/publications/Miscellaneous,
  Accessed 27 Dec 2022.

\bibitem[Wen et~al., 2022]{wen2022impacts}
Wen, J., Wang, S., Yang, X., and Zhou, X. (2022).
\newblock Impacts of epidemics on innovation: An empirical analysis.
\newblock {\em Technovation}, page 102634.

\bibitem[Yamashita, 2021]{yamashita2021economic}
Yamashita, N. (2021).
\newblock Economic crisis and innovation capacity of japan: Evidence from
  cross-country patent citations.
\newblock {\em Technovation}, 101:102208.

\bibitem[Yoon, 2021]{Yoon21}
Yoon, C. (2021).
\newblock How does covid-19 affect digital innovation and transformation?
\newblock {\em KDI School of Pub Policy \& Management Paper SSRN}, (4054290).

\bibitem[Zerhouni et~al., 2020]{Zerhounietal20}
Zerhouni, W., Nabel, G.~J., and Zerhouni, E. (2020).
\newblock Patents, economics, and pandemics.
\newblock {\em Science}, 368(6495):1035--1035.

\end{thebibliography}

\clearpage
\appendix
\section{Age at patent publication}
\label{app:data}
\begin{figure}[H]
    \begin{center}
 \includegraphics[width=.7\textwidth]{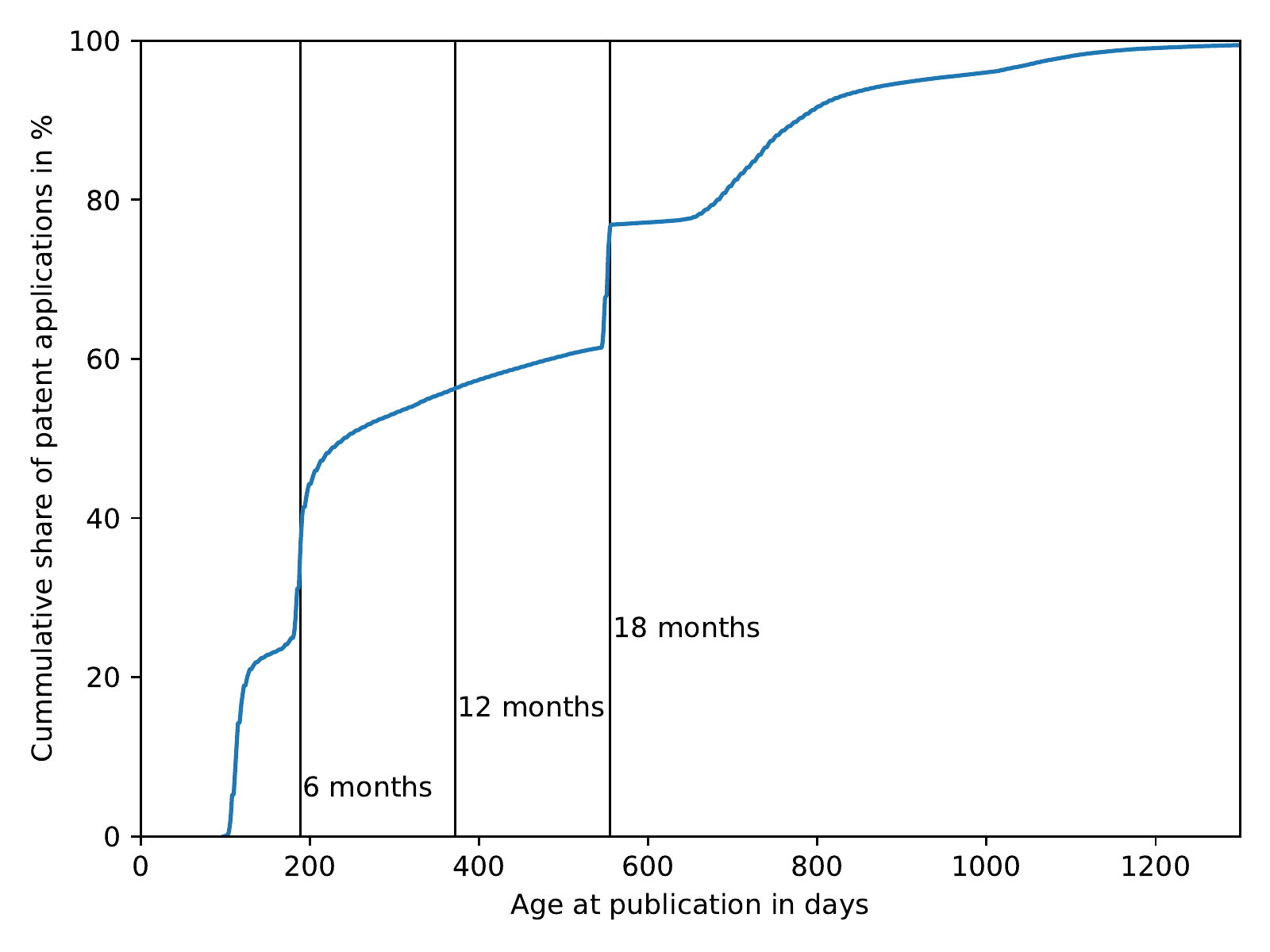}
\caption{Cumulative distribution of age at publication for patent applications published 2014-2022. A majority of 76.67\% is published within 18 months.}
\label{fig:age-at-publication}  
    \end{center}
\end{figure}

Figure~\ref{fig:age-at-publication} shows the cumulative distribution of age at publication for patent applications published 2014-2022. More than three-quarters of patents are published within 18 months of filing. 

\section{OLS residual plot}
\label{app:ols_res}
\begin{figure}[H]
\begin{center}
	\begin{minipage}{.49\textwidth}
		\includegraphics[width=\textwidth]{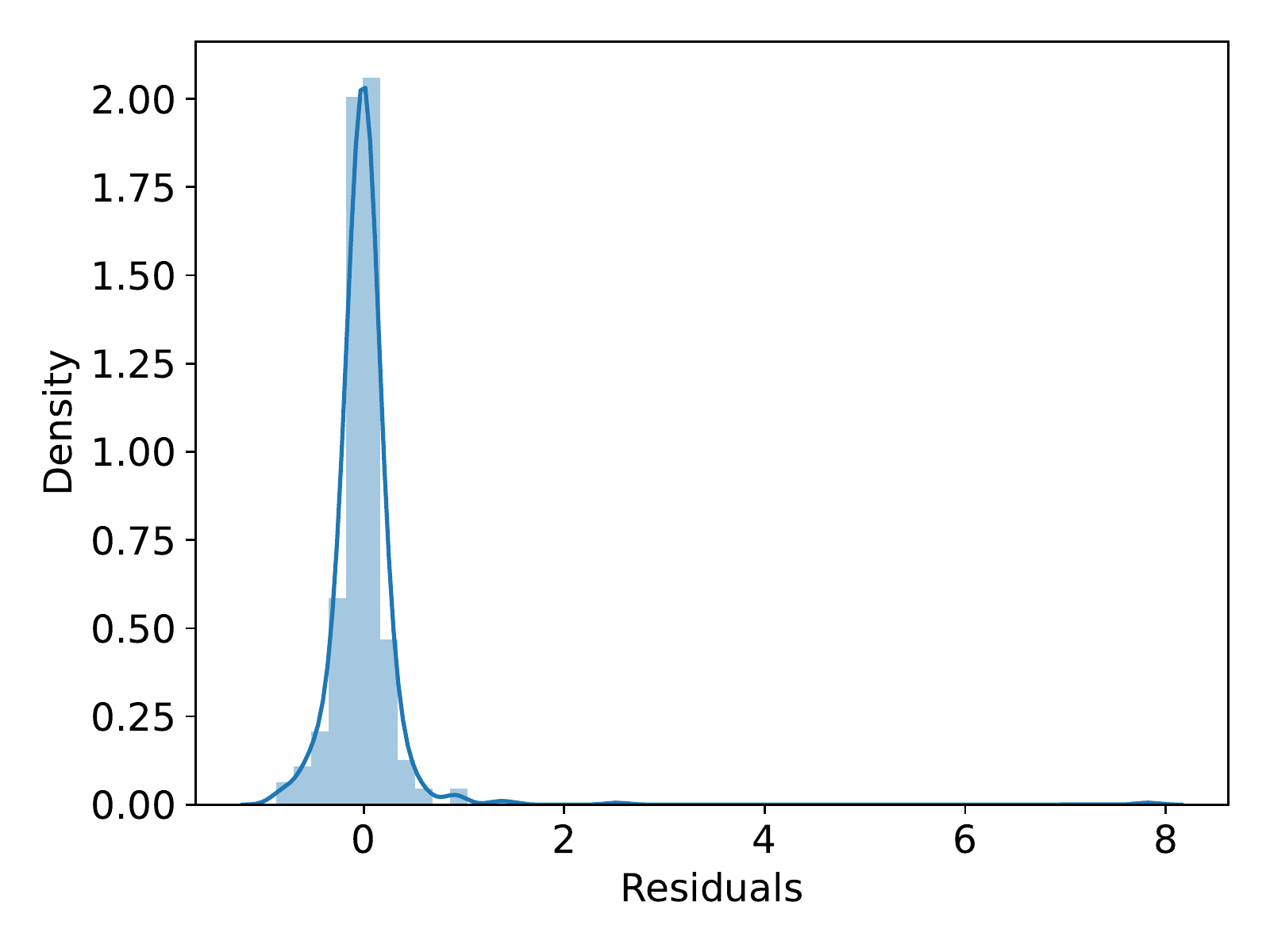}
	\end{minipage}
		\begin{minipage}{.49\textwidth}
		\includegraphics[width=\textwidth]{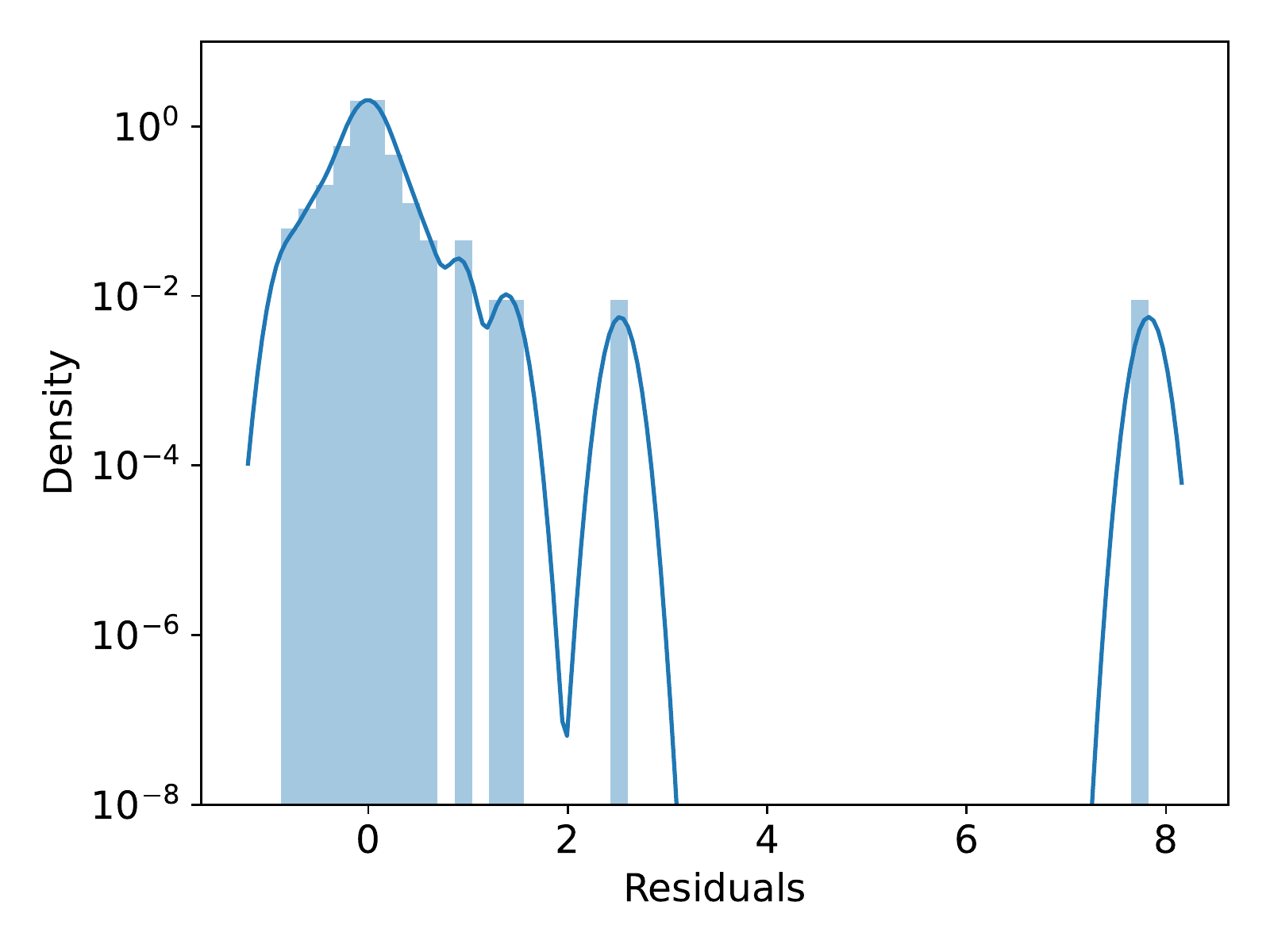}
	\end{minipage}
 \caption{Distribution of OLS residuals in model (7); linear (left) and logarithmic scale (right).}
\label{fig:residuals:lin}   
\end{center}
\end{figure}

Figure~ \ref{fig:residuals:lin} shows the distribution of the residuals in model (7) in table~\ref{tab:ols:g19}. The distribution exhibits the non-normality with high skewness and kurtosis, suggesting that the OLS fit is  most likely invalid and possibly misleading.

%
%

%

\section{Directly and indirectly most Covid-19 relevant CPC subclasses}
\label{app:c19relevance}

To give an indication of the makeup of the variables $CovidRelevant$ and $IndirectCovidRelevant$, we show the 50 subclasses with the highest values for each in tables~\ref{tab:C19direct} (direct Covid-19 relevance) and \ref{tab:C19indirect} (indirect Covid-19 relevance). For most entries (albeit not for all) it is immediately obvious why this field of technology would be directly/indirectly relevant for the management of the pandemic situation. It is also clear that the table for indirect Covid-19 relevance contains fields more associated with basic technologies, e.g. from classes C07 (organic chemistry), C08 (organic macromolecular compounds) G06 (computing, calculating, counting), H01 (basic electric elements), H04 (electric communication technique), etc.

Table~\ref{tab:mainsections} lists the descriptions of the nine CPC mainsections.

\begin{table}[tbh!]
\centering
\begin{tabular}{p{2.5cm} p{11cm}}
\hline\hline
\textbf{CPC \mbox{Mainsection}} & \textbf{Description} \\
\hline\hline
A & Human necessities \\
B & Performing operations, transporting \\
C & Chemistry, metallurgy \\
D & Textiles, paper \\
E & Fixed constructions \\
F & Mechanical engineering, lighting, heating, weapons, blasting \\
G & Physics \\
H & Electricity \\
Y & New technology tags \\
\hline\hline
\end{tabular}
    \caption{CPC (Cooperative Patent Classification System) mainsections}
    \label{tab:mainsections}
\end{table}

\begin{table}[p!]
\vspace*{-1.5cm}
\hspace*{-0.5cm}\begin{tabular}{p{1.75cm} p{14cm} p{1.75cm}}
\hline\hline
\textbf{CPC \mbox{Subclass}} & \textbf{Description} & \textbf{Share in Percent} \\
\hline\hline
A61P & Specific therapeutic activity of chemical compounds or medicinal preparations & 0.605 \\
Y02A & Technologies for adaptation to climate change & 0.411 \\
A61K & Preparations for medical, dental, or toilet purposes & 0.379 \\
A62B & Devices, apparatus or methods for life-saving & 0.370 \\
C12Q & Measuring, testing processes involving enzymes, nucleic acids, microorganisms, etc. & 0.353 \\
A41D & Outerwear; protective garments; accessories & 0.326 \\
G16B & Bioinformatics & 0.289 \\
C12N & Microorganisms or enzymes; compositions thereof; genetic engineering; etc. & 0.287 \\
A01N & Preservation of bodies of humans, animals, plants; biocides, etc. & 0.260 \\
C07K & Peptides & 0.252 \\
G16H & Healthcare informatics & 0.239 \\
B01L & Chemical or physical laboratory apparatus for general use & 0.181 \\
G16Z & Information and comm. technology adapted for other specific application fields& 0.179 \\
A61L & Methods or apparatus for sterilising materials or objects & 0.176 \\
C07D & Heterocyclic compounds & 0.144 \\
G01N & Investigating materials by determining chemical or physical properties & 0.142 \\
G16C & Computational chemistry; chemoinformatics; computational materials science & 0.138 \\
C07H & Sugars; derivatives thereof; nucleosides; nucleotides; nucleic acids & 0.128 \\
F24F & Air-conditioning; ventilation; etc.; use of air currents for screening & 0.097 \\
C12R & Indexing scheme associated with C12C-Q, relating to microorganisms & 0.095 \\
C40B & Combinatorial chemistry; libraries, e.g. chemical libraries & 0.088 \\
A61H & Physical therapy apparatus & 0.086 \\
A23Y & Indexing scheme relating to lactic etc. bacteria used in food & 0.083 \\
C11D & Detergent compositions; soap; resin soaps; recovery of glycerol; etc. & 0.082 \\
D06M & Treatment of fibres, threads, yarns, etc. not provided for elsewhere in D06 & 0.080 \\
B82Y & Specific uses, applications, measurement, manufacture of nanostructures & 0.076 \\
A61M & Devices for introducing media into, or onto, the body, etc. & 0.073 \\
A61N & Electrotherapy; magnetotherapy; radiation therapy; etc. & 0.073 \\
C12Y & Enzymes & 0.073 \\
A23K & Fodder & 0.072 \\
B03C & Magnetic or electrostatic separation of solid materials & 0.068 \\
C12M & Apparatus for enzymology or microbiology & 0.067 \\
A21D & Treatment, e.g. preservation, of flour or dough & 0.048 \\
C05D & Inorganic fertilisers not covered by C05B-C; Fertilisers producing CO2 & 0.046 \\
B60H & Arrangements of heating, cooling, ventilating passenger spaces, etc. & 0.041 \\
G16Y & Information and comm. technology specially adapted for the internet of things & 0.040 \\
A23B & Preserving, e.g. by canning, meat, etc.; chemical ripening, etc. & 0.037 \\
H05H & Plasma technique (other electric techniques) & 0.037 \\
A01K & Animal husbandry & 0.037 \\
B01J & Chemical or physical processes, e.g. catalysis & 0.037 \\
C12P & Fermentation or enzyme-using processes to synthesise a chemical compound, etc. & 0.035 \\
A61B & Diagnosis; surgery; identification & 0.034 \\
A61G & Transport, personal conveyances, etc. for patients or disabled persons & 0.034 \\
A24D & Cigars; cigarettes; tobacco smoke filters; etc. & 0.032 \\
A45C & Purses; luggage; hand carried bags & 0.030 \\
A23L & Foods, etc. not covered by A21D, A23B-J; Preservation of foods & 0.028 \\
B01D & (Chemical) separation & 0.028 \\
E05B & Locks; accessories therefor; handcuffs & 0.028 \\
C07B & General methods of organic chemistry; apparatus therefor & 0.027 \\
A47L & Domestic washing or cleaning; suction cleaners & 0.026 \\
\hline\hline
\end{tabular}
    \caption{50 top CPC subclasses in terms of direct Covid-19 relevance}
    \label{tab:C19direct}
\end{table}

\begin{table}[p!]
\vspace*{-1.5cm}
\hspace*{-0.5cm}\begin{tabular}{p{1.75cm} p{14cm} p{1.75cm}}
\hline\hline
\textbf{CPC \mbox{Subclass}} & \textbf{Description} & \textbf{Share in Percent} \\
\hline\hline
Y10T & Technical subjects covered by former US classification & 0.380 \\
G06F & Electric digital data processing & 0.296 \\
Y10S & Subjects covered by former USPC cross-reference art collections and digests & 0.256 \\
H01L & Semiconductor devices; electric solid state devices not otherwise provided for & 0.243 \\
H04L & Transmission of digital information, e.g. telegraphic communication & 0.181 \\
H04N & Pictorial communication, e.g. television & 0.141 \\
A61K & Preparations for medical, dental, or toilet purposes & 0.129 \\
A61B & Diagnosis; surgery; identification & 0.107 \\
G01N & Investigating materials by determining chemical or physical properties & 0.104 \\
G02B & Optical elements, systems or apparatus & 0.091 \\
C07D & Heterocyclic compounds & 0.089 \\
G11B & Information storage based on relative movement between record carrier and transducer & 0.081 \\
H04W & Wireless communication networks & 0.080 \\
G06Q & Data processing systems or methods, for administrative, commercial, etc. purposes & 0.073 \\
C07C & Acyclic or carbocyclic compounds & 0.067 \\
C08L & Compositions of macromolecular compounds & 0.060 \\
H04B & Transmission (electric communication) & 0.059 \\
H05K & Printed circuits; casings or constructional details of electric apparatus; etc. & 0.058 \\
G11C & Static stores (information storage)& 0.058 \\
B29C & Shaping or joining of plastics, etc. & 0.057 \\
B01D & (Chemical) separation & 0.053 \\
G06K & Graphical data reading; presentation of data; record carriers; etc. & 0.052 \\
C07K & Peptides & 0.052 \\
A61F & Filters implantable into blood vessels, etc. & 0.051 \\
Y02T & Climate change mitigation technologies related to transportation & 0.051 \\
H04M & Telephonic communication & 0.051 \\
G01R & Measuring electric or magnetic variables & 0.049 \\
H01R & Electrically-conductive connections; coupling devices; current collectors; etc. & 0.049 \\
A61M & Devices for introducing media into, or onto, the body, etc. & 0.048 \\
B01J & Chemical or physical processes, e.g. catalysis & 0.047 \\
G06T & Image data processing or generation, in general & 0.047 \\
B65D & Containers for storage or transport of articles or materials & 0.047 \\
C12N & Microorganisms or enzymes; compositions thereof; genetic engineering; etc. & 0.047 \\
Y02E & Reduction of greenhouse gas emissions, related to energy & 0.046 \\
Y02P & Climate change mitigation technologies in production etc. & 0.046 \\
C08G & Macromolecular compounds obtained not by reactions of only unsatur. C-to-C bonds & 0.045 \\
G03G & Electrography; electrophotography; magnetography & 0.044 \\
B41J & Typewriters; selective printing mechanisms, etc. & 0.039 \\
H01J & Electric discharge tubes or discharge lamps & 0.039 \\
G02F & Optical devices for the control of light; non-linear optics; etc. & 0.038 \\
B60R & Vehicles, vehicle fittings, or vehicle parts, not otherwise provided for & 0.036 \\
E21B & Earth drilling, e.g. deep drilling; obtaining oil, gas, water, etc. & 0.035 \\
C08F & Macromolecular compounds obtained by reactions of only c-to-c unsaturated bonds & 0.034 \\
H01M & Direct conversion of chemical energy into electrical energy, e.g. batteries & 0.034 \\
H03K & Pulse technique (electronic circuitry) & 0.033 \\
G03F & Photomechanical production of textured or patterned surfaces & 0.033 \\
G01S & Radio direction-finding; radio navigation; etc. & 0.032 \\
A63B & Apparatus for physical training, etc. & 0.032 \\
C23C & Coating of or with metallic material; surface treatment by chemical conversion, etc. & 0.032 \\
G09G & Arrangements or circuits for control of indicating devices & 0.032 \\
\hline\hline
\end{tabular}
    \caption{50 top CPC subclasses in terms of indirect Covid-19 relevance}
    \label{tab:C19indirect}
\end{table}

\end{document}